\pdfoutput=1

\documentclass[reprint,aps,prb,twocolumn,floatfix,superscriptaddress]{revtex4-2}

\usepackage[linktocpage=true,colorlinks=true,urlcolor=blue,linkcolor=red,citecolor = blue]{hyperref}
\usepackage{graphicx}
\usepackage{amssymb}
\usepackage{bm}
\usepackage{amsmath}
\usepackage{amsfonts}
\usepackage{dsfont}
\usepackage{xcolor}
\usepackage{lineno}
\usepackage[english]{babel}
\usepackage[utf8]{inputenc}
\RequirePackage[T1]{fontenc}
\RequirePackage{lmodern}
\usepackage{appendix}
\usepackage{float}
\usepackage{algpseudocode}
\usepackage{algorithm}
\usepackage{mathtools}
\usepackage{physics}
\usepackage{comment}
\usepackage{siunitx}
\usepackage{bbm}
\usepackage{soul}
\begin{document}

\title{The importance of shear strain in narrow Si/SiGe quantum wells}

\author{Hudaiba Soomro}
\affiliation{University of Wisconsin-Madison, Madison, WI 53706 USA}

\author{Talise Oh}
\affiliation{University of Wisconsin-Madison, Madison, WI 53706 USA}

\author{M. A. Eriksson}
\affiliation{University of Wisconsin-Madison, Madison, WI 53706 USA}

\author{Benjamin D. Woods}
\affiliation{University of Wisconsin-Madison, Madison, WI 53706 USA}

\author{Mark Friesen}
\affiliation{University of Wisconsin-Madison, Madison, WI 53706 USA}

\date{\today}

\begin{abstract} 
Several heterostructure profiles have been proposed for Si/SiGe quantum wells, to improve the operation of quantum-dot spin qubits. 
These structures share a common goal of generating either of the two wave vectors that govern the conduction-band valley coupling, dubbed $k_0$ and $k_1$. 
Here, we focus on the long-wavelength $k_1$ term that couples valleys when paired with shear strain, and which governs the Dresselhaus spin–orbit coupling. 
We show that shear strain induced by metal top gates may be enough to activate the $k_1$ valley-coupling mechanism, mitigating the need for ``ultrasharp'' interfaces, as commonly assumed for large valley energy splittings.
We propose a scheme featuring narrow quantum wells that does not require ultrasharp interfaces, but should still provide large deterministic valley splittings (when paired with shear strain), and large spin-orbit couplings and $g$-factor corrections, which are of great interest for experimental implementations.
\end{abstract}

\maketitle

\begin{figure*}[t] 
	\includegraphics[width=12cm]{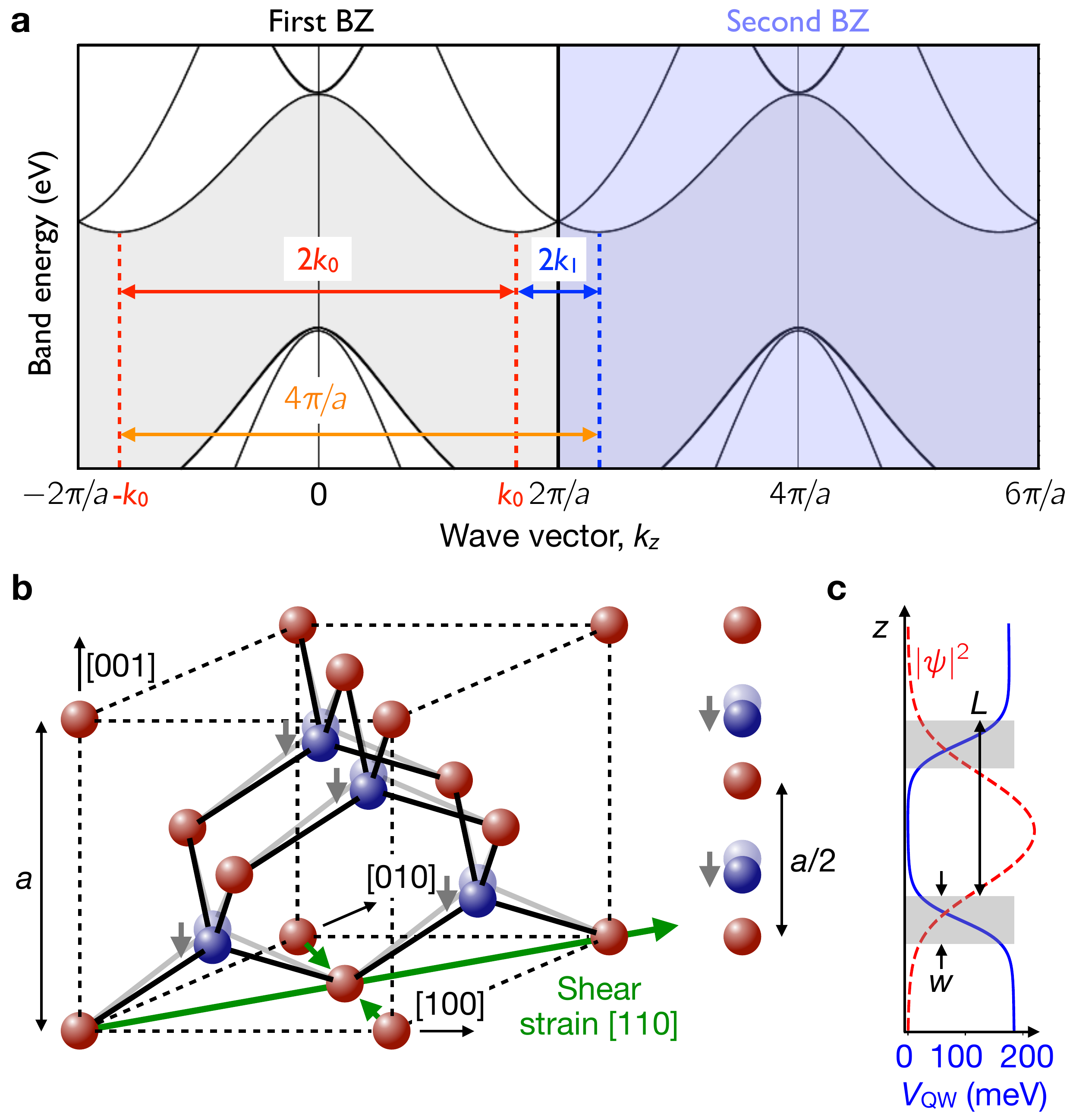}
	\centering
	\caption{
    \textbf{Using shear strain and small feature sizes to enhance the valley splitting.}
    \textbf{a} Valley coupling requires a wave vector of length $2k_0$ (red arrow), as illustrated in the band structure of the first Brillouin zone (BZ) for Si.
    Shear strain provides an alternative (second-order) scheme for coupling the valleys, by combining a lattice deformation of wave vector $k_\text{shear}=4\pi/a$ (orange arrow) with an engineered feature in the quantum well confinement potential $V_\text{QW}(z)$ of wave vector $2k_1=k_\text{shear}-2k_0$.
    Here, $a$ is the size of the Si cubic unit cell, and the required feature size is $\lambda_1=2\pi/2k_1\approx 1.60$~nm.
    \textbf{b} The Si cubic unit cell comprises a diamond lattice, with its two sublattices highlighted in red and blue.
    Shear strain can be achieved by stretching along [110] and squeezing along $[1\bar{1}0]$ (green arrows).
    As a result, the blue sublattice shifts toward $[00\bar{1}]$ (gray arrows), as shown on the right, yielding a lattice deformation of period $\lambda_\text{shear}=a/2$, with a corresponding reciprocal wave vector of $k_\text{shear}=4\pi/a$.
    \textbf{c} In this work, we consider a narrow quantum well defined by the sigmoid function of Eq.~(\ref{eq:ge_profile}), with a quantum well width $L$ and/or an interface width $w$ comparable to $\lambda_1$.
    Here, we show an example wave function (red dashed line) computed for a quantum well defined by $L=15~\text{ML}= 2.04~\text{nm}$ and $w=4~\text{ML}= 0.54~\text{nm}$, in a quantum well with a band offset of 180~meV and a vertical electric field of 2~mV/nm.
    Other types of engineered features in $V_\text{QW}(z)$ could also provide the required $2k_1$ wave vector, including a Wiggle Well or a simple ``bump.''}
	\label{fig:cartoon}
\end{figure*}

Recent years have witnessed the rapid expansion of quantum dots in large one- \cite{Zajac:2018p439, Zwerver:2022p184, Nickl:2026p5878} and two-dimensional spin-qubit arrays \cite{Unseld:2023p084002, Ha:2025p030327, John:2025p10560}, by leveraging advances in the semiconductor industry. 
Underlying this progress is the ability to operate devices at uniform voltage settings \cite{Neyens2024p80},
easing the burden of tune-up and large-scale operation.
Despite these successes, materials variability remains a key challenge across solid-state qubit platforms \cite{deLeon:2021peabb2823}.
For quantum dots formed in silicon-germanium quantum wells, the random nature of the SiGe alloy is a persistent and fundamental source of variability \cite{Wuetz:2022p7730, Losert:2023p125405}, as reflected in local fluctuations of energy splittings between the conduction-band valley eigenstates \cite{Boykin:2004p115, Friesen:2007p115318, Volmer:2024p61}, their complex phases \cite{woods2025preprint, king2026preprint}, the spin-orbit coupling \cite{Nestoklon:2006p235334, Prada:2011p013009, Ferdous:2018p26, Cai:2023p386}, and the Land\'e $g$-factor \cite{Woods:2026p206201, volmer2026preprint}.

Historically, the main focus of valley splitting experiments has been on extracting mean values \cite{Burkard:2023p025003}. 
However, recent theory work highlights the role of alloy disorder and the importance of characterizing valley splitting \emph{distributions}, which are known to take a Rician form in the most general case \cite{Losert:2023p125405, Thayil:2025p115303}.
In recent years, systematic measurements of the valley splitting (e.g., across dot arrays or shuttlers \cite{Volmer:2024p61}, or in devices where dots are continuously positioned between plunger gates \cite{Marcks:2025p11381}) point to disorder-dominated behavior, which is well-described by the limiting Rayleigh distribution.
Consequently, a significant fraction of dots will have low valley splittings, rendering such devices inoperable, as the spin states hybridize with excited valley states lying outside the logical subspace.

To overcome these challenges, several schemes have been proposed to deterministically enhance the valley splitting \cite{Losert:2023p125405}.
Some recent experiments also point to possible deterministic enhancement \cite{Abraham:2026p1154, Esposti2026}, although the source of such behavior is unclear. 
On the other hand, all confinement schemes experience disorder-induced fluctuations of the valley splitting, due to the overlap of electron wavefunctions with disordered SiGe barriers. In fact, some schemes for enhancing the valley splitting require a significant Ge overlap. At the same time, the disorder also leads to spatially varying potential-energy landscapes that affect the shape and location of a dot, the tunnel coupling between dots, and even the formation of spurious dots. Optimal solutions to the valley splitting problem should therefore weigh the potential benefits from deterministic enhancement against complications arising from disorder.

The most straightforward scheme for coupling valleys requires engineering a quantum well confinement potential $V_\text{qw}(z)$ with a strong Fourier component at the special wave vector, $k_z=2k_0$, that directly connects the $z$ conduction valleys on opposite sides of the Brillouin zone \cite{Friesen:2007p115318}, as illustrated in Fig.~\ref{fig:cartoon}a.
Here, $k_0\approx 0.83(2\pi/a)$, where $a=0.543$~nm is the cubic lattice constant of Si, shown in Fig.~\ref{fig:cartoon}b.
Heterostructures capable of providing such large wave vectors include the short-period Wiggle Well \cite{McJunkin:2022p7777, Feng:2022p085304} and quantum wells with ultrasharp interfaces \cite{Boykin:2004p115, Chen:2021p044033}.
Unfortunately, the characteristic size of these features, $\lambda_0=2\pi/2k_0=0.33$~nm, is so small that they are challenging to obtain by standard industrial techniques like chemical vapor deposition (CVD) \cite{Koch:2025p59}.
(For recent progress based on molecular beam epitaxy, see \cite{rahlff2026preprint}.)

What then is the source of the persistently high valley splittings observed in some experiments? 
A recent proposal suggests that the strict requirement of ultrasharp ($2k_0$) features can be relaxed in the presence of shear strain \cite{Woods2024}. 
In this case, a much smaller wave vector $2k_1$ is needed to couple the valleys, via the second-order process shown in Fig.~\ref{fig:cartoon}a (and explained below).
The characteristic feature size for this coupling, $\lambda_1=2\pi/2k_1=1.60$~nm, is fully compatible with current CVD growth techniques \cite{McJunkin:2022p7777}.

In this work, we show that narrow quantum wells can be used to deterministically enhance the valley splitting in the presence of shear strain. 
We also show that the quantum dot devices recently reported in \cite{Abraham:2026p1154} are well-suited for inducing shear strain at the levels needed for generating consistently high valley splittings, and we speculate that devices with micromagnets could further enhance this effect.
In fact, we find that narrow quantum wells with shear strain can successfully compete with Wiggle Wells, which were specifically designed for this purpose \cite{Woods2024}.
We then show that the $\sim$3~nm quantum wells used in recent experiments \cite{Chen:2021p044033, Acuna:2024p044057} could activate a combination of $k_0$ and $k_1$-based enhancement mechanisms, and we conclude that, since such strains are present, they should certainly be taken into account. 
Finally, we note that the same conditions that enable high valley splittings can also enhance the Dresselhaus spin-orbit coupling and the Land\'e $g$-factor.

\begin{figure*}[t] 
	\includegraphics[width=13cm]{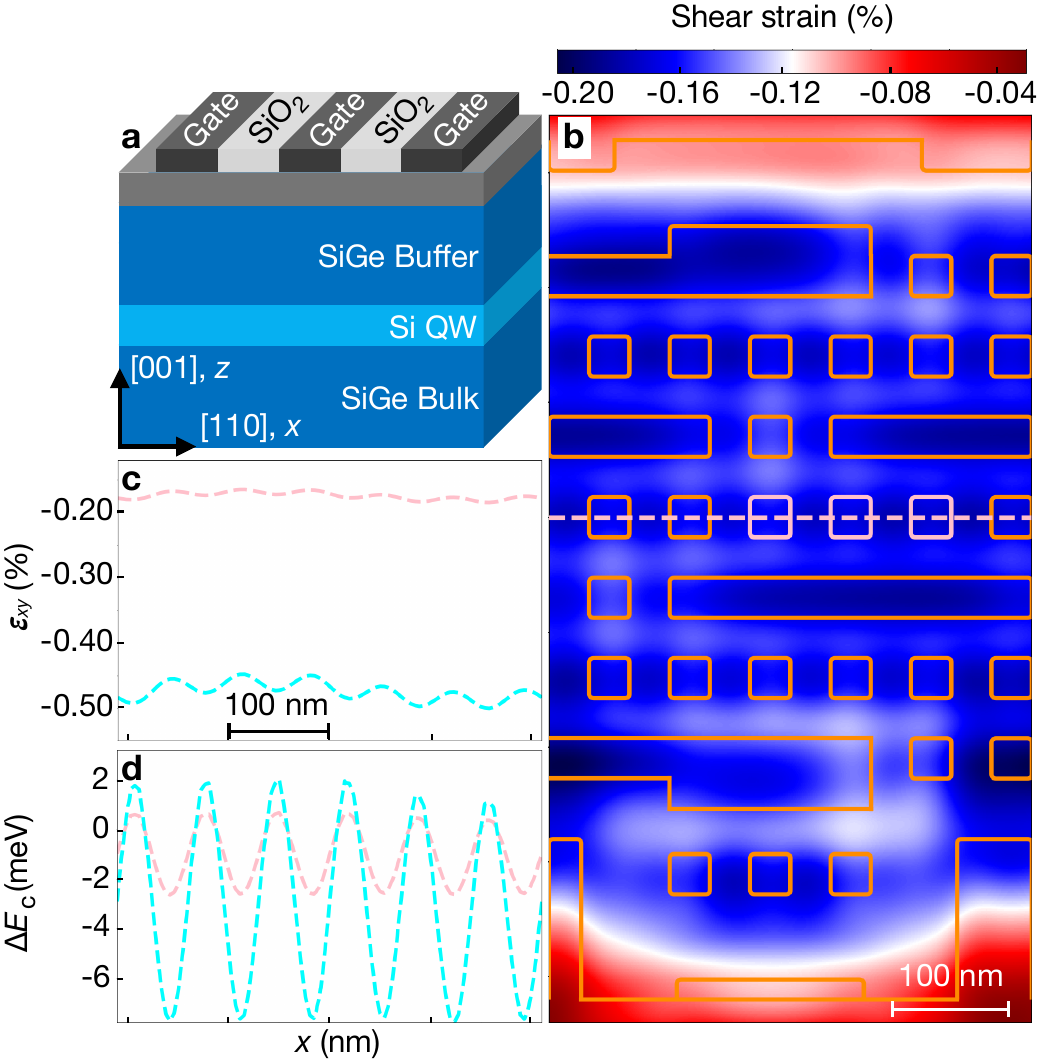}
	\centering
	\caption{
    \textbf{Shear strain simulations. 
    a} A cross-section view of the simulated device in the $x$-$z$ plane (not drawn to scale), where $\hat{x}$ is defined along the [110] direction. 
    From bottom to top, the simulations include a Si$_{0.7}$Ge$_{0.3}$ virtual substrate (dark blue), a Si quantum well (light blue), a Si$_{0.7}$Ge$_{0.3}$ spacer layer (dark blue), an insulating Al$_2$O$_3$ layer (medium gray), and W metal gates (dark gray), surrounded by SiO$_{2}$ oxide (light gray), with corresponding thicknesses of 2~$\mu$m, 3~nm, 30~nm, 5~nm, 300~nm, and 300~nm. 
    \textbf{b} A simulated shear-strain color map for the gate geometry outlined in orange.
    (The three gates outlined in pink are referenced in other figures, below.)
    The gate structure is based on \cite{Abraham:2026p1154}. 
    As in \cite{Abraham:2026p1154}, the gate pitch is 70~nm, while the spacing between the gates is chosen to be 35~nm. 
    The barrier gates outlined on the top and bottom of the figure are assumed to extend to the distant edges of the simulated device (see Methods and Supplementary Note~S1).  
    \textbf{c} The dashed pink curve shows a linecut of the shear strain in \textbf{b}, evaluated along the horizontal dashed line. 
    The dashed blue curve shows a similar linecut obtained from Supplementary Fig.~S1d.
    These two results are obtained for different depositional stress values for W, taken from the literature (see Methods). 
    \textbf{d} Strain-induced conduction-band energy fluctuations for the same linecuts shown in \textbf{c}. 
    For visual clarity, these curves are shifted vertically, so the regions between the gates have an average value of $\Delta E_c=0$.
    }
	\label{fig:strain}
\end{figure*}

\section*{Results}
\subsection*{Shear strain}
We recall that shear strain corresponds to the off-diagonal component of the strain tensor, $\varepsilon_{xy}$.
As an example, we consider a lattice distortion oriented along the crystallographic [110] axis, whose effect on the Si diamond crystal lattice is illustrated in Fig.~\ref{fig:cartoon}b.
Here, the shear strain is introduced by stretching along the [110] axis while squeezing along $[1\bar{1}0]$, as indicated by the green arrows.
The effect of these stresses is to displace the diamond crystal sublattices (indicated by blue and red) in opposite directions, causing a bilayer pairing of the atomic planes along [001] (gray arrows), as illustrated on the right-hand side of the panel.
The resulting periodicity along $\hat z$ is $\lambda_\text{shear}=a/2$, with a corresponding wave vector of $k_\text{shear}=4\pi/ a$, which couples points across the Brillouin zone, as indicated in Fig.~\ref{fig:cartoon}a.
A second-order process involving the wave vector $2k_1$ then completes an effective $2k_0$ coupling between the two valleys.
(See \cite{Woods2024} for additional details.)

\begin{figure*}[t] 
	\includegraphics[width=14cm]{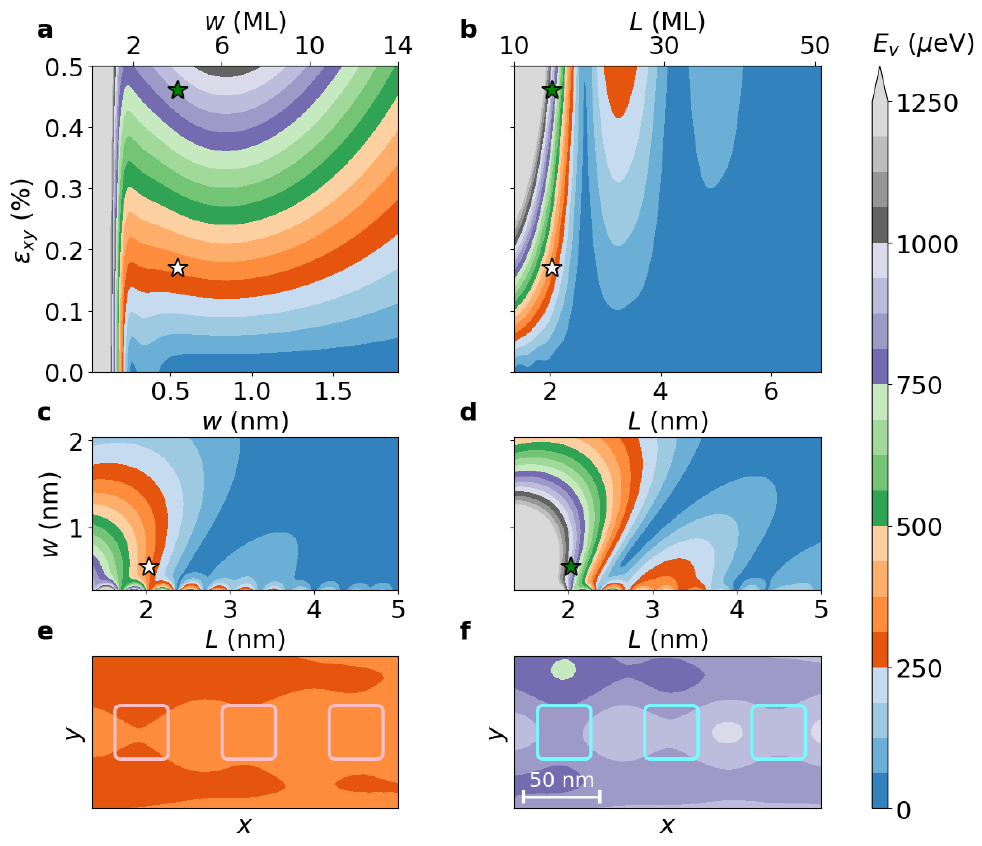}
	\centering
	\caption{
    \textbf{Deterministic valley splitting results, based on effective-mass simulations, in the absence of random-alloy disorder. 
    a} Color map of the valley splitting $E_v$, as a function of the interface width $w$ and shear strain $\varepsilon_{xy}$, for a fixed quantum-well width of $L=15~\text{ML}=2.04$~nm. 
    ($w$ is indicated in units of nm and ML.) 
    In all panels, the stars identify the parameters $(w,L)=(0.54~\text{nm},2.04~\text{nm})$ for two cases: $\varepsilon_{xy} = 0.17\%$ (white star) and $\varepsilon_{xy} = 0.46\%$ (green star), corresponding to the average shear strains beneath the plunger gates, for the pink and blue curves in Fig.~\ref{fig:strain}c.
    \textbf{b} Color map of the valley splitting as a function of $L$ and $\varepsilon_{xy}$, for the fixed value $w=4~\text{ML}=0.54$~nm, revealing an interference pattern governed by the valley-coupling wave vector $2k_1$. 
    ($L$ is indicated in units of nm and ML.)
    \textbf{c, d} Color maps of the valley splitting as a function of $L$ and $w$, for fixed values of \textbf{c} $\varepsilon_{xy}=0.17\%$, or \textbf{d} $0.46\%$. 
    (Also see Supplementary Fig.~S3.)
    \textbf{e, f} Spatial maps of the deterministic valley splitting for the fixed values $L=15~\mathrm{ML}$ and $w=4~\mathrm{ML}$, using shear-strain maps from \textbf{e} Fig.~\ref{fig:strain}b, or \textbf{f} Supplementary Fig.~S1d. 
    Here, the colors of the gates are the same as the highlighted gates in those figures. }
	\label{fig:valley_splitting}
\end{figure*}

An important consideration for the $2k_1$ valley-coupling mechanism is the direction of the applied stress.
Although shear strain can be defined in any coordinate system, valley coupling requires shear strain defined with respect to the crystallographic axes of the cubic unit cell, as illustrated in Fig.~\ref{fig:cartoon}b.
Such strain occurs naturally in many device settings, due to the presence of defects in the crystal lattice, including cross-hatch, whose features are naturally oriented along [110] and $[1\bar{1}0]$ \cite{Schaffler:1991p2039}.
Shear strain can also be engineered: for example, in \cite{Woods2024}, etched trenches were found to cause shear strain, while in \cite{Emily2026preprint}, shear strain was generated by stressors fabricated on top of the gates.
In \cite{frink2025preprint}, very large shear strains of order 0.4\% were observed below a fabricated micromagnet. 

In this work, we focus on strains arising from metal electrodes \cite{Thorbeck:2015p087107}, noting that significant shear-strain components can be achieved when the devices are aligned with respect to [110] or $[1\bar{1}0]$.
As an important example, we consider the device described in \cite{Abraham:2026p1154}, which exhibits an unexpectedly high valley splitting.
This industrially fabricated device incorporates gate structures in the form of vertical vias, composed predominantly of W.
Such tall gate structures are expected to enhance the strain, as compared to the shorter, planar gates, commonly preferred in academic settings.
Additionally, some of the gates have high aspect ratios, which we also expect to enhance the shear strain \cite{Emily2026preprint}.
Since detailed specifications for the heterostructure and the gate stack in \cite{Abraham:2026p1154} have not been publicly disclosed, we adopt the details that have been provided, while filling in the missing information with parameters reported for older devices \cite{Chen:2021p044033, Acuna:2024p044057}. 
A cross-sectional view of the device assumed in our simulations is shown in Fig.~\ref{fig:strain}a (not drawn to scale), with parameters described in the caption and Methods.

Typical results from the shear-strain simulations are presented in Fig.~\ref{fig:strain}b.
Here, the $\hat x$ direction is defined as the characteristic axis of the device geometry, which we assume to be along [110]. 
In this work, we consider two different values for the W depositional stress as reported in the literature (see Methods).
Fig.~\ref{fig:strain}b shows results for the more moderate stress value of 1.12~GPa~\cite{Kim:2000p577}, while Supplementary Fig.~S1b also shows results for a larger stress value of 4~GPa~\cite{LEE2026166754}.
(Fig.~S1 also shows results for a device oriented along [100].)
The gate structures, based on \cite{Abraham:2026p1154}, are outlined in orange or pink in Fig.~\ref{fig:strain}b.
In Fig.~\ref{fig:strain}c, we present a more detailed view of the simulation results, for both depositional stress values, evaluated along the linecut shown in Fig.~\ref{fig:strain}b (horizontal dashed line).
A key result of these simulations is that the absolute value of the shear strain can range from 0.15-0.5\% under every plunger gate, depending on the metal deposition method used. 
Comparing these results to the imposed stress conditions, we infer that depositional stress is the dominant source of shear strain for the blue curve, but that thermal contractions play a significant role for the pink curve.
(See \cite{frink2025preprint} for a related discussion.)
We note that the strain approaches zero, as expected, far away from this central region of the device (Supplementary Fig.~S1).
We also note that the overall trends of the shear strain can be explained by arguments like those presented in \cite{Emily2026preprint}, where the ``outer field gates’’ at the top and bottom of Fig.~\ref{fig:strain}b play the same role as the stressors in \cite{Emily2026preprint}.

In the discussion that follows, we show that the strain levels in Fig.~\ref{fig:strain}b and Supplementary Fig.~S1 can provide consistently large valley splittings, depending on the details of the heterostructure.
In Fig.~\ref{fig:strain}d, we also report the potential energy fluctuations caused by the diagonal elements of the strain tensor, which strongly modify the confinement potential over distances comparable to the gate pitch.
These fluctuations can have deep minima below the plunger gates and large barriers between the plungers, suggesting that strong electrostatic potentials could be needed to overcome these effects.
In contrast, the shear-strain variations shown in Fig.~\ref{fig:strain}c are relatively smooth below the plunger gates. 

\subsection*{Deterministic valley splitting}

We now compute the valley splitting using effective-mass methods.
The theory, expressed in Eq.~(\ref{eq:valley_coupling_operator}) of Methods, includes both $2k_0$ and $2k_1$ terms, but allows these effects to be explored independently.
As discussed above, valley splitting is sensitive to the shape of the quantum well confinement potential $V_\text{qw}(z)$, especially features containing $2k_0$ or $2k_1$ wave vectors.
Here, we assume sigmoidal confinement potentials (see Methods), characterized by the quantum-well width $L$ and the interface width $w$, as illustrated in Fig.~\ref{fig:cartoon}c.
The sigmoidal shape is physically realistic \cite{Dyck:2017p1700622} and, importantly, it avoids introducing sharp corners into the confinement potential, which artificially generate $2k_0$ and $2k_1$ wave vector components \cite{Losert:2023p125405}.
In the simulations below, we consider a range of shear-strain values, $0\leq\varepsilon_{xy}\leq 0.5$\%, as consistent with the strain simulations reported in Fig.~\ref{fig:strain}.
We note that the results in this section do not include disorder, and are therefore \emph{deterministic}.
Disorder effects will be introduced in a later section.

In Figs.~\ref{fig:valley_splitting}a-d, we present valley splitting results as a function of $\varepsilon_{xy}$, $w$, and $L$, while holding a different parameter constant in each plot.
Throughout this work, we also assume a uniform, vertical electric field of $F_z=2$~mV/nm, which forces electrons against the top quantum-well interface.
In Fig.~\ref{fig:valley_splitting}a, we set $L=15~\text{ML}= 2.04$~nm, where ML refers to the thickness of a single atomic monolayer ($1~\text{ML}\approx 0.136$~nm).
On the far-left side of the map, we observe high valley splittings with no dependence on shear strain; this behavior arises from the direct $2k_0$ valley coupling caused by ultrasharp interfaces with $w\lesssim \lambda_0=0.33$~nm.
In the rest of the map, we observe $2k_1$-dominated behavior, which depends approximately linearly on $\varepsilon_{xy}$.
The two regimes are separated by a knee-like feature at $w\approx \lambda_0$.
In Supplementary Note~S2, we confirm this crossover interpretation, from $2k_0$ to $2k_1$-dominated behavior, by turning these terms on and off in the Hamiltonian.
Next, in Fig.~\ref{fig:valley_splitting}b, we set $w=4~\text{ML}= 0.54$~nm, which does \emph{not} qualify as an ultrasharp interface; this map, therefore, does not exhibit any substantial $2k_0$-type behavior. 
However, it does display interference fringes occurring when the wave function interacts with both the top and bottom quantum-well interfaces, analogous to interference fringes observed previously in the $2k_0$ regime \cite{Boykin:2004p115}.
In Fig.~\ref{fig:valley_splitting}b, the fringes are separated by $\Delta L\approx 1.6$~nm -- the characteristic length scale associated with $2k_1$-type behavior.
Finally, in Figs.~\ref{fig:valley_splitting}c and d, we fix the values of $\varepsilon_{xy}$: in Fig.~\ref{fig:valley_splitting}c we set $\varepsilon_{xy}=0.17$\%, corresponding to the pink curve in Fig.~\ref{fig:strain}c, while in Fig.~\ref{fig:valley_splitting}d, we set $\varepsilon_{xy}=0.46$\%, corresponding to the blue curve in Fig.~\ref{fig:strain}c. 
In both cases, we observe very strong enhancements of the valley splitting on the lower-left region of the map, induced by the $2k_1$ mechanism, when either $w$ or $L$ approaches 1.6~nm.
We also observe a series of interference fringes at the bottom of both maps, corresponding to $2k_0$ interference effects; in this case, the distance between the fringes is closer to 0.16~nm, characteristic length scale for $2k_0$-type behavior.
As a reference, in Figs.~\ref{fig:valley_splitting}c and d, we highlight the quantum-well parameters $(w,L)=(0.54~\text{nm},2.04~\text{nm})$ with stars, corresponding to the deterministic valley splittings $0.33$ and $0.88$~meV, respectively.
The same star locations are also shown in Figs.~\ref{fig:valley_splitting}a and b.

\begin{figure*}[t] 
	\includegraphics[width=14cm]{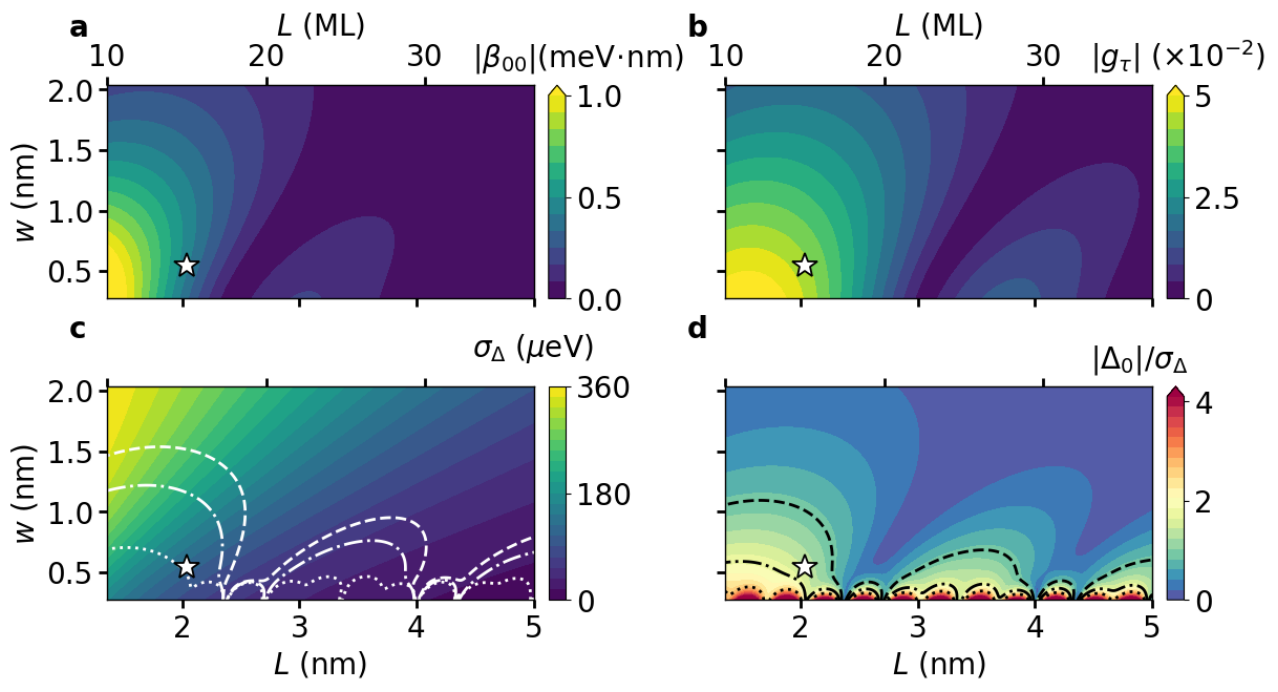}
	\centering
	\caption{
    \textbf{Dependence of the spin-orbit coupling, $g$-factor, and disorder parameters on the shape of the quantum well. 
    a} Color map of the Dresselhaus spin–orbit coupling strength $\lvert\beta_{00}\rvert$, plotted as a function of quantum-well parameters $L$ and $w$. 
    \textbf{b} Color map of the $g$-factor corrections $\lvert g_\tau\rvert$ (see Methods), as a function of $L$ and $w$. 
    \textbf{c} Color map of the standard deviation of the valley coupling $\sigma_\Delta$, as a function of $L$ and $w$. Dashed, dot-dashed, and dotted lines indicated the nominal crossover to deterministically enhanced valley splitting behavior ($|\Delta_0|/\sigma_\Delta = \sqrt{\pi}/2$), for the cases $\varepsilon_{xy} = 0.1, 0.2$, and 0.3\%, respectively.
    \textbf{d} Color map of the deterministic-crossover ratio $|\Delta_0|/\sigma_\Delta$, as a function of $L$ and $w$, for the fixed value $\varepsilon_{xy} =0.17$\%. 
    Dotted, dot-dashed, and dashed lines correspond to the contours $|\Delta_0|/\sigma_\Delta = \sqrt{\pi}/2, \sqrt{\pi}$, and $3\sqrt{\pi}/2$, respectively.
    (Also see Supplementary Fig.~S5.)}
	\label{fig:SOC_g_tau}
\end{figure*}

The stars in Figs.~\ref{fig:valley_splitting}a-d represent desirable working points, owing to their large valley splitting values.
Figure~\ref{fig:valley_splitting}b reveals additional $2k_1$ oscillation fringes that could serve as working points.
For example, the second fringe is centered at a convenient quantum-well width of $L\approx 3.4$~nm, which has already been achieved in recent experiments.
The $2k_1$ fringes are also observed in Fig.~\ref{fig:valley_splitting}, and particularly, in Fig.~\ref{fig:valley_splitting}d, where the $2k_0$ and $2k_1$ fringes work together to give higher valley splitting values.
The simulations therefore reveal a potential interplay between $2k_0$ and $2k_1$ phenomena that is experimentally accessible.

Figures~\ref{fig:valley_splitting}e and f show spatial maps of the deterministic valley splitting, computed with the strain profiles of Fig.~\ref{fig:strain} and Supplementary Fig.~S1, for the three gates highlighted in each of those figures.
Since the simulations in this section do not include disorder, the variations observed here are attributed strictly to gate-induced strain variations.
Summarizing these results: we have demonstrated the presence of consistently large valley splittings in the $2k_1$ regime, which are accessible to current experiments.
For sharper interfaces, $2k_0$ behavior may also emerge, further enhancing the valley splitting.

\begin{figure*}[t] 
	\includegraphics[width=12.3cm]{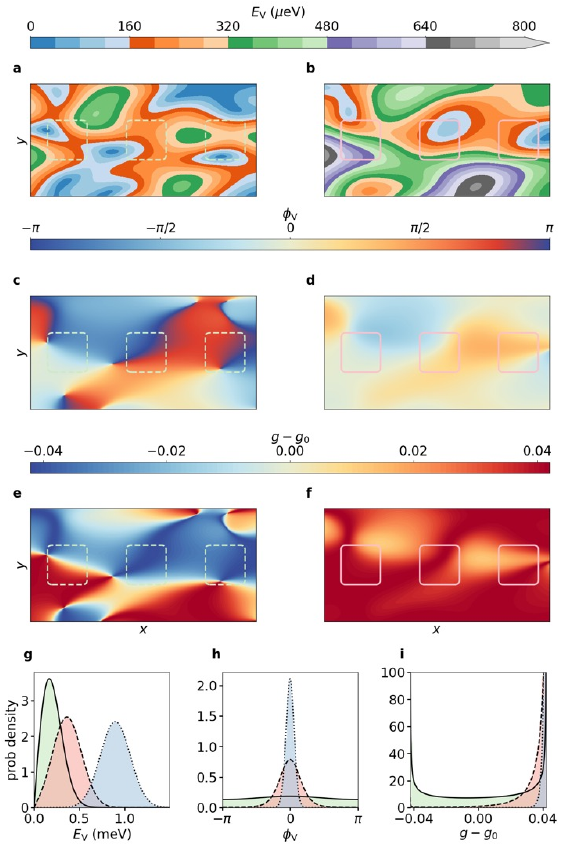}
	\centering
	\caption{
    \textbf{Comparison of disorder-dominated vs deterministically enhanced behaviors. 
    a-f} Color maps on the left are obtained with shear strain turned off, $\varepsilon_{xy}=0$, while maps on the right are obtained using the strains computed in Fig.~\ref{fig:strain}b.
    All maps are computed using the same random-alloy disorder realization.
    The outlined gates refer to the same outlined gates as Figs.~\ref{fig:valley_splitting}e and f: on the right-hand side, the pink gates generate the same shear strains as the pink gates in  Fig.~\ref{fig:valley_splitting}e; however, on the left-hand side, the dashed green outlines are only shown for reference, since we set $\varepsilon_{xy}=0$.
    We assume a sigmoid quantum-well profile, defined by $L=15~\mathrm{ML}$ and $w=4~\mathrm{ML}$. 
    \textbf{a, b} Valley-splitting maps showing \textbf{a} disorder-dominated behavior, or \textbf{b} deterministic enhancement.
    \textbf{c, d} Valley-phase maps showing \textbf{c} disorder-dominated behavior, or \textbf{d} deterministic enhancement.
    Note that valley vortices are observed in \textbf{c} but not in \textbf{d}.
    \textbf{e, f} Maps of the fluctuating $g$-factor, $g-g_0$, showing \textbf{e} disorder-dominated behavior, or \textbf{f} deterministic enhancement.
    \textbf{g-i} Sample-wide distributions of \textbf{g} the valley splitting, \textbf{h} the valley phase, and \textbf{i} the $g$-factor fluctuations, obtained from $10^7$ independently sampled disorder realizations, assuming spatially uniform $\varepsilon_{xy}$ values: green corresponds to $\varepsilon_{xy}=0$ (disorder-dominated), pink corresponds to $\varepsilon_{xy}=0.17$\% (deterministically enhanced), blue corresponds to  $\varepsilon_{xy}=0.46$\% (strongly enhanced).
    Each distribution is overlaid with its theoretical probability distribution function (solid, dashed, and dotted lines; see Supplementary Note~S5).
    In \textbf{g}, deterministic enhancement causes a suppression of low valley splitting values, while in \textbf{h} and \textbf{i}, it causes a narrowing of the distributions.}
	\label{fig:disorder}
\end{figure*}

\subsection*{Spin-orbit coupling and \texorpdfstring{$g$}{g}-factor}

The dominant spin-orbit coupling in the Si/SiGe system takes the Dresselhaus form, proportional to $(\hat{k}_x\sigma_x - \hat{k}_y\sigma_y)$, where $\hat{k}_j=-i\partial_j$, and $\sigma_j$ are Pauli matrices acting on the electron spin.
Since this coupling is an intervalley process \cite{Woods:2023p035418}, it shares some common features with the valley splitting. 
Unlike the valley splitting, however, the Dresselhaus coupling has a built-in wave vector of $4\pi/a$, and therefore does not require any additional shear strain \cite{Woods:2023p035418}.
In fact, the Dresselhaus coupling mechanism is described by the same second-order process illustrated in Fig.~\ref{fig:cartoon}a; so, it also needs a confinement potential with a $2k_1$ wave vector. 
As discussed in Methods, the strength of the Dresselhaus coupling in the ground subband is defined as $\beta_{00} = \beta_0 \mel{\varphi_0}{e^{i2k_1z}}{\varphi_0}$, where $\beta_0=8.2~\mathrm{meV\,nm}$ is the bare Dresselhaus coupling coefficient for Si. 
The ground subband envelope function, $\varphi_0(z)$, is the same envelope used in valley splitting calculations; it contains spatial fluctuations that can induce spin-orbit coupling if they contain a $2k_1$ wave vector, inherited from the quantum well confinement potential $V_\text{qw}(z)$.

Figure~\ref{fig:SOC_g_tau}a shows results for $|\beta_{00}|$ as a function of $w$ and $L$.
As expected, we observe large enhancements for narrow quantum wells and sharp interfaces.
For example, at the starred location in Fig.~\ref{fig:SOC_g_tau}a (the same starred location as Fig.~\ref{fig:valley_splitting}c), we obtain $|\beta_{00}|\approx 0.43$~meV$\cdot$nm.
We can compare this to a conventional quantum well (e.g., with $w=1$~nm and $L=10$~nm), for which we find $|\beta_{00}|\approx 6.7$~$\mu$eV$\cdot$nm, representing an enhancement factor of 65 for the narrow well.
Due to their shared intervalley origins, the spin-orbit coupling shown in Fig.~\ref{fig:SOC_g_tau}a resembles the valley splitting distribution in Fig.~\ref{fig:valley_splitting}c, except for the absence of $2k_0$ features at low $w$, since the $2k_0$ wave vector plays no (direct) role in the spin-orbit coupling.
The results also do not depend on the shear strain, to leading order.
(At higher order, small strain and $k_0$-based corrections can occur, since these parameters enter the Hamiltonian and affect $\varphi_0$.)
In Supplementary Note~S3, we show that $E_v\propto|\beta_{00}|$ in the $2k_1$-dominated regime, to a very good approximation.

The spin-orbit coupling also contributes to the Land\'e $g$-factor \cite{Woods:2026p206201}, resulting in a correction to the bare coupling $g_0\approx 2$, proportional to $|g_\tau|\cos (\phi_v-\phi_g)$ (see Methods).
Here, $|g_\tau|$ is the magnitude of the correction, while $\phi_g$ is its complex phase and $\phi_v$ is the phase of the intervalley coupling (known as the ``valley phase’’).
In Fig.~\ref{fig:SOC_g_tau}b, we plot $|g_\tau|$ as a function of $w$ and $L$, also showing strong enhancement when these quantities approach 1.6~nm.
As in Fig.~\ref{fig:SOC_g_tau}a, we observe no $\varepsilon_{xy}$ or $k_0$ dependence here, to leading order.
For more-conventional quantum wells, corresponding to larger $w$ and $L$ values, $|g_\tau|$ is of order $2\times10^{-3}$ \cite{volmer2026preprint}.
However, the $2k_1$ enhancement mechanism studied here yields much larger values; for example, at the starred location in Fig.~\ref{fig:SOC_g_tau}b, we obtain $|g_\tau|\approx 4.2 \times 10^{-2}$.

\subsection*{Spatial variations due to the SiGe random alloy}

Due to their small size, electrons in quantum dots are exposed to a limited number of Ge atoms within a Si/SiGe quantum well.
In some structures, the Ge atoms are concentrated in the barrier regions, where they mainly contact the tails of the electron wave functions \cite{Wuetz:2022p7730}.
In other cases, Ge is intentionally added to the quantum well, where the electron density is highest, in the form of spikes \cite{McJunkin:2021p085406}, Wiggle Wells \cite{McJunkin:2022p7777, rahlff2026preprint}, or a uniform doping layer \cite{Marcks:2025p11381}.
Since the SiGe alloy is random, the Ge atoms are located at random sites throughout the SiGe crystal lattice, causing the electrons to sample random, local variations of the Ge density.
Moreover, since Ge and Si atoms have different on-site potentials and bond lengths, this variability also leads to spatial fluctuations of the quantum dot confinement potential, valley splitting, and other local properties \cite{Losert:2023p125405, Volmer:2024p61, volmer2026preprint, soomro2026preprint}.
The spatial variations of the intervalley coupling are known to follow a Gaussian distribution characterized by its standard deviation $\sigma_\Delta$ \cite{Losert:2023p125405}, which depends on the laterally-averaged Ge spatial profile $n_\text{Ge}(z)\in (0,1)$ of the quantum well, in the form of a weighted average: $\sigma_\Delta^2\propto l_t^{-2}\int dz\, |\varphi_0(z)|^4 n_\text{Ge}(z)[1-n_\text{Ge}(z)]$.
Throughout this work, we set $l_t \approx 20~\text{nm}$ as the in-plane confinement length for a quantum dot, corresponding to a typical orbital confinement energy of $\hbar\omega_t = 1$ meV.
The spatial variations of the intravalley coupling (i.e., the potential energy landscape) also follow a Gaussian distribution with a standard deviation of $\sqrt{2}\sigma_\Delta$ \cite{Nemeth:2026p010336}.
In Fig.~\ref{fig:SOC_g_tau}c, we plot $\sigma_\Delta$ as a function of the quantum-well parameters $w$ and $L$.
This confirms our expectation that smaller $L$ and larger $w$ should both cause the electron wave function to overlap with more Ge, thereby increasing the alloy disorder felt by the electron, and $\sigma_\Delta$.
Indeed, we find that $\sigma_\Delta$ depends roughly linearly on both $w$ and $L$.

Previously, in Fig.~\ref{fig:valley_splitting}, we studied the deterministic component of the valley splitting, obtained in the virtual crystal approximation, which does not include alloy disorder.
Here, we explore this disorder and the emergence of deterministically enhanced valley splitting.
We consider the same quantum-well profile and strain maps as Figs.~\ref{fig:valley_splitting}e and f.
However, we now include intervalley disorder potentials using the statistical methods of Refs.~\cite{Losert:2023p125405, Woods:2026p206201} (see Supplementary Note~S4).
In Fig.~\ref{fig:disorder}a, we first show a spatial map of the valley splitting for $(L,w)=(15,4)$~ML, including disorder but no shear strain -- in this case, $2k_1$ effects are not activated.
(Compare to Figs.~\ref{fig:valley_splitting}e and f, which include strain effects but no disorder.)
We note that $L$ and $w$ are also not small enough to activate $2k_0$ behavior; indeed, the $2k_0$ contribution to the deterministic valley splitting is only $23.4~\mu$eV here (see Supplementary Fig. S2a).
In this case, since the quantum well is narrow, the electron wave function overlaps with many Ge atoms, endowing the valley splitting with a large standard deviation ($\sigma_\Delta=118$~$\mu$eV) and a large disorder-enhanced mean value ($\bar{E_v}=0.21$~meV).
However, many dot locations still have low valley splittings, as indicated by dark blue in Fig.~\ref{fig:disorder}a.
Figure~\ref{fig:disorder}b shows a second valley splitting map, with the same disorder profile as Fig.~\ref{fig:disorder}a, but now including the shear strain from Fig.~\ref{fig:strain}b.
While these results still appear disordered, the average valley coupling has been significantly enhanced by the $2k_1$ coupling mechanism.
In particular, the regions of very low valley splitting are suppressed.
We also note that the gate structure is not clearly visible in the valley splitting map, since the shear strain is largely uniform across this window.

The crossover to deterministically enhanced behavior depends on two key parameters \cite{Losert:2023p125405}: the deterministic valley coupling, $\Delta_0$, representing the center of the valley-coupling distribution in the complex plane, and the standard deviation of this distribution, $\sigma_\Delta$.
We note that $\Delta_0$ is also the valley coupling that would be observed in the absence of alloy disorder.
In turn, these quantities are determined by the quantum-well confinement profile $V_\text{qw}(z)$, which is averaged across the heterostructure, and by the vertical electric field $F_z$ and shear strain $\varepsilon_{xy}$, which are locally varying.
The term \emph{deterministically enhanced valley splitting} refers to cases where the valley-coupling distribution is centered far enough from zero that dangerously small valley splittings are appreciably suppressed.
The crossover to this regime occurs when \cite{Losert:2023p125405} $|\Delta_0|/\sigma_\Delta\gtrsim\sqrt{\pi}/2\approx 0.9$, although strong suppression of small valley-splitting values occurs when this ratio takes a larger value, in the range of $|\Delta_0|/\sigma_\Delta\gtrsim (2\text{-}3)\times \sqrt{\pi}/2\approx 1.8$-2.7 (see Supplementary Fig.~S7).

In Fig.~\ref{fig:SOC_g_tau}d, we plot the crossover ratio $|\Delta_0|/\sigma_\Delta$ as a function of the quantum-well parameters $w$ and $L$, revealing interference patterns reminiscent of Fig.~\ref{fig:valley_splitting}c, with two distinct length scales arising from the different $2k_0$ and $2k_1$ valley-coupling mechanisms. 
The dashed black lines in Fig.~\ref{fig:SOC_g_tau}d highlight several crossover criteria for the deterministic enhancement, from ``normal'' enhancement to very strong enhancement; by comparing these curves to Fig.~\ref{fig:valley_splitting}c, we note that deterministic enhancement can occur even in cases where the valley splitting is not especially high ($\sim$100~$\mu$eV). 
The white dashed lines in Fig.~\ref{fig:SOC_g_tau}c also indicate contours of deterministic enhancement -- in this case, revealing that significant disorder can accompany high valley splittings.
The valley-splitting maps in Figs.~\ref{fig:disorder}a and b exemplify both sides of the deterministic crossover, with $|\Delta_0|/\sigma_\Delta=0.1$ (Fig.~\ref{fig:disorder}a) corresponding to ``randomly dominated'' behavior and $|\Delta_0|/\sigma_\Delta=1.4$  (Fig.~\ref{fig:disorder}b) corresponding to ``deterministically enhanced'' behavior.
These results confirm that the $2k_1$ coupling mechanism can boost the valley splitting well into the deterministically enhanced regime.
The sample-wide valley-splitting distributions for these different cases are shown in Fig.~\ref{fig:disorder}g
(also see Supplementary Figs.~S7a-d).
In particular, for the shear-strain level corresponding to the blue line in Fig.~\ref{fig:strain}c, we observe a very strong suppression of low valley-splitting values.
Since $|\Delta_0|\propto\varepsilon_{xy}$ for $2k_1$-based valley coupling, this suggests a clear path for improving the valley splitting: by increasing the shear strain.

The valley phase $\phi_v$ also provides a key signature of deterministic behavior \cite{woods2025preprint}, although measuring it requires special experimental techniques \cite{volmer2026preprint, king2026preprint}.
In Figs.~\ref{fig:disorder}c and d, we plot spatial maps of the valley phase, obtained for the same device parameters and disorder profiles as the valley-splitting maps in Figs.~\ref{fig:disorder}a and b.
For the case of no shear strain (Fig.~\ref{fig:disorder}c), we observe a full range of valley-phase values, $\phi_v\in[0,2\pi]$, consistent with the center of the complex $\Delta$ distribution lying near the origin.
In Fig.~\ref{fig:disorder}d, on the other hand, we observe a much narrower range of valley phases, consistent with the $\Delta$ distribution shifting away from the origin.
The sample-wide distribution functions for these valley-phase maps are shown in Fig.~\ref{fig:disorder}h, for the same three parameter profiles as Fig.~\ref{fig:disorder}g.
(Note that the valley phases here and in Fig.~\ref{fig:disorder}c and d have been reported relative to $\phi_g$.)
It is clear that the narrowing of the $\phi_v$ distributions provides a definitive signature of deterministic enhancement \cite{woods2025preprint}.
For example, while the pink $E_v$ distribution in Fig.~\ref{fig:disorder}g looks similar to the randomly dominated distribution in green, the pink $\phi_v$ distribution in Fig.~\ref{fig:disorder}h shows clear deterministic enhancement.

Valley-phase spatial maps like Figs.~\ref{fig:disorder}c and d also display other interesting features.
For example, the deterministically enhanced map (Fig.~\ref{fig:disorder}d) contains no valley vortices, which can be identified as singularities with $E_v=0$, around which the phase winds by $2\pi$.
However, we contrast this with the disorder-dominated map (Fig.~\ref{fig:disorder}c), which is obtained from the same disorder profile, but exhibits four vortices across the same window.
The difference between these behaviors is also a consequence of deterministic enhancement, reflecting the suppression of valley-splitting zeros.

Finally, we note that Ge concentration fluctuations are atomistic by definition, giving rise to $2k_0$ effects, which have a nearly atomistic length scale of 0.16~nm.
However, such fluctuations are largely smoothed out over the much larger $2k_1$ length scale of  1.60~nm.
Consequently, quantities like the spin-orbit coupling ($\beta_{00}$) and $g_\tau$, which are enhanced by $k_1$ features (not $k_0$), are largely robust to atomistic Ge fluctuations.
In contrast, $g$-factor fluctuations arise from the quantity $|g_\tau|\cos (\phi_v-\phi_g)$, and are therefore susceptible to Ge fluctuations through the valley phase $\phi_v$, resulting in spatial variations. 
In Figs.~\ref{fig:disorder}e and f, we plot $g$-factor maps for the same parameter profiles as the valley-splitting and valley-phase maps in Figs.~\ref{fig:disorder}a-d.
These results mirror the valley phase map.
For example, we observe a significant narrowing of the $g$-factor distribution for the deterministically enhanced case (Fig.~\ref{fig:disorder}f), as compared to the disorder-dominated case (Fig.~\ref{fig:disorder}e).
Also analogous to Figs.~\ref{fig:disorder}c and d, we observe valley vortices in Fig.~\ref{fig:disorder}e, but not Fig.~\ref{fig:disorder}f.
The narrowing of the distribution is more fully illustrated in Fig.~\ref{fig:disorder}i, where we plot sample-wide distributions of the $g$-factor, obtained for the same three parameter profiles as Figs.~\ref{fig:disorder}g and h.

\section*{Discussion}

In summary, we have shown that a long-held assumption -- that ``ultrasharp'' quantum-well interfaces (defined to be narrower than $\lambda_0=0.33$~nm) are needed to obtain large, deterministic valley splittings -- can be relaxed in the presence of shear strain.
In this scenario, a much larger (and experimentally favorable) value of $\lambda_1=2\pi/2k_1=1.60$~nm emerges as the characteristic length scale for valley coupling.
Using a combination of simulation techniques, we have shown that gate geometries featured in recent experiments \cite{Abraham:2026p1154} could provide enough shear strain to generate large deterministic enhancements of the valley splitting, when coupled with sufficiently narrow quantum wells \cite{Chen:2021p044033, Acuna:2024p044057}, and that these same heterostructures should strongly enhance the Dresselhaus spin-orbit coupling and $g$-factor fluctuations.
We also speculate that wider quantum wells with internal features like steps, wiggles, or spikes could provide the required $2k_1$ wave vector, and that micromagnets could alternatively provide the needed shear strain \cite{frink2025preprint}.

We conclude with some final remarks on $2k_1$-based coupling.
First, we emphasize that the relatively large $\lambda_1$ length scale is essentially immune to atomic-scale disorder at the quantum-well interface.
This should be contrasted with the $2k_0$ coupling mechanism, which is highly sensitive to single-atom steps \cite{Losert:2023p125405}.
Second, as strain engineering has only recently emerged as an active topic for quantum-dot qubits, further experiments are needed to characterize and exploit its effects.
Indeed, our results indicate that shear strain could be present in many device settings, and may also arise from anomalous sources, such as cross-hatch phenomena.
Here, we suggest that narrow quantum wells could provide a convenient platform for harnessing these effects.
Conversely, we argue that experiments encountering shear strain should account for possible $2k_1$ effects when analyzing energy dispersions.
Finally, we emphasize that experiments hosting high valley splittings via the $2k_1$ mechanism should also experience enhanced Dresselhaus spin-orbit couplings and $g$-factor fluctuations, since these phenomena are governed by the same $2k_1$ wave vector.

\section*{Acknowledgements}
The authors are grateful for many enlightening discussions with E. S. Joseph.
We also acknowledge discussions with L.M.K. Vandersypen and the S. Tarucha group, which motivated and clarified the importance of micromagnets for generating shear strain in these devices.

\section*{Funding}
This research was sponsored in part by the Army Research Office under Award No.\ W911NF-23-1-0110.
T.O.\ was sponsored by the Netherlands Ministry of Defence under Award No.\ QuBits R23/009. 
The views, conclusions, and recommendations contained in this document are those of the authors and are not necessarily endorsed by nor should they be interpreted as representing the official policies, either expressed or implied, of the Army Research Office, the U.S. Government, or the Netherlands Ministry of Defence. The U.S. Government is authorized to reproduce and distribute reprints for U.S. Government purposes notwithstanding any copyright notation herein. The Netherlands Ministry of Defence is also authorized to reproduce and distribute reprints for Government purposes notwithstanding any copyright notation herein.

\section*{Competing interests}
M.A.E., B.D.W., and M.F.\ have applied for a patent on strained Si/SiGe quantum wells. 
The remaining authors declare no competing interests.

\section*{Methods}
\subsection*{Shear strain modeling in COMSOL}
\noindent The strain modeling in this work closely follows the Methods described in \cite{frink2025preprint}, which we summarize as follows.

Simulations are performed using the Solid Mechanics module of COMSOL Multiphysics~\cite{Comsol}. 
The simulations require input values for the Young's modulus $E$, the Poisson's ratio $\nu$, and the coefficients of thermal expansion (CTE) $\alpha$, for each material. 
While all parameters depend on temperature, the dependence of $E$ and $\nu$ is relatively weak, so we adopt room-temperature values for these quantities. 
We define temperature-averaged CTEs as $\bar{\alpha} = (T_i - T_f)^{-1}\int_{T_f}^{T_i} \alpha(T)\, dT$, where $T_i$ and $T_f$ are the initial and final temperatures of the cool-down, respectively. 
These integrals are approximated using data provided in \cite{bradley2013, Simon1994, McSkimin1953, Levinshtein2000HandbookInSb, Corruccini1961ThermalLiterature, Burke1962ThermalC}. 
We use Vegard's law to evaluate the CTE of the SiGe alloy. 
The resulting temperature-averaged CTEs are given by $\{\bar{\alpha}_{\text{SiO}_{2}}, \bar{\alpha}_{\text{Si}}, \bar{\alpha}_{\text{Si}_{0.7}\text{Ge}_{0.3}}, \bar{\alpha}_{\text{W}}, \bar{\alpha}_{\text{Al}}\} = \{0.5, 0.76, 1.52, 4.5, 14.16\}$ $(\times 10^{-6} \text{~K}^{-1})$. 
We use the full, orthotropic compliance tensors for Si and SiGe materials in our COMSOL simulations, as detailed in \cite{frink2025preprint}. 
Biaxial strain for Si in the quantum well, arising from lattice mismatch with a strain-relaxed Si$_{0.7}$Ge$_{0.3}$ barrier, is given by $\varepsilon_{xx} = \varepsilon_{yy} = 1.2707\%$ and $\varepsilon_{zz} = -0.9798\%$. 
This strain is implemented in COMSOL as an initial-strain parameter. 

Depositional stress from the fabrication of metal gates is included as an initial-stress parameter that accounts for the effects of thermal contraction as the metal is cooled from depositional temperatures to $293~\mathrm{K}$. 
Consistent with the experiments described in \cite{Abraham:2026p1154}, we adopt W metal gates for our simulations.
We specifically consider two different values of the depositional stress for these W gates: $1.12 ~\mathrm{GPa}$ \cite{Kim:2000p577} and $4 ~\mathrm{GPa}$ \cite{LEE2026166754}, to account for possible variations in fabrication methods.
No depositional stress is assumed for Al$_2$O$_3$ due to the wide range of reported values, including sign changes, and their sensitivity to deposition conditions and thermal post-processing \cite{Ylivaara2022ThermomechanicalDeposition}. We do not expect this approximation to qualitatively affect the results, but it may cause small quantitative errors. 
For the same reason we do not include any depositional stress for SiO$_2$.

Free boundary conditions~\cite{Comsol} are enforced on every surface, except for the bottom, where we enforce a contraction that matches the bulk thermal contraction of Si, which has the effect of emulating the presence of a much larger Si substrate at this boundary. 
The final results of our strain simulations are evaluated in the plane of the two-dimensional electron gas, assumed to lie $1.5~\text{nm}$ below the top interface of the quantum well, where the low-energy electronic wave functions of the quantum well are concentrated. 
All device dimensions have been thoroughly checked, to avoid finite-size effects in the region of interest.

In the main text, we consider the gate structure whose outlines are shown in Fig.~\ref{fig:strain}b, oriented along the [110] crystallographic axis, while in Supplementary Note~S1, we also consider a device oriented along [100].
To simplify the simulations in the [110] orientation, we consider a reference frame defined by $\hat{x}'=[110]$, $\hat{y}'=[\bar{1}10]$, and $\hat{z}'=\hat{z}=[001]$.
In this frame, the device is oriented along $\hat x'$.
However, the orthotropic compliance tensors must also be rotated by $45^\circ$, as described in \cite{frink2025preprint}.
As noted in the main text, the key input parameter for valley-coupling calculations is the shear strain evaluated in the \emph{crystallographic} coordinate system.
Thus, we must perform an inverse transformation between these two frames as the final step of the simulation procedure.
The desired shear strain is then given by
\begin{equation}
   \varepsilon_{xy} = (\varepsilon_{x'x'} - \varepsilon_{y'y'})/2 .
\end{equation}

In addition to valley-splitting energy variations arising from the local shear strain, we also observe variations of the conduction-band offsets $\Delta E_c$ arising from the diagonal strain components.
(For example, see Fig.~\ref{fig:strain}d.)
We compute these variations here using the standard relation~\cite{Thorbeck:2015p087107} 
\begin{equation}
    \Delta E_c = E_0 + \Xi_u \varepsilon_{zz} + \Xi_d (\varepsilon_{xx} + \varepsilon_{yy} + \varepsilon_{zz}), \label{DeltaEc}
\end{equation}
where $\Xi_u = 10.5$~eV and $\Xi_d = 1.1$~eV are deformation-potential coefficients~\cite{Fischetti1996BandAlloys}. 
We note that Eq.~(\ref{DeltaEc}) is independent of the coordinate system: $(x,y,z)\leftrightarrow(x',y',z')$.
Unless otherwise stated, $E_0$ is shifted in each calculation, such that the asymptotic value of $\Delta E_c$ goes to zero, far away from the active region containing the plunger gates.

\subsection*{Effective-mass theory}
In this work, we use a one-dimensional effective-mass model to calculate the valley splitting $E_v$, the intrasubband (ground-state) Dresselhaus spin-orbit coupling coefficient $|\beta_{00}|$, and the magnitude of the intervalley $g$-factor renormalization parameter $|g_\tau|$. 
This model is derived from a multiorbital $sp^{3}d^{5}s^{*}$ tight-binding theory in Refs.~\cite{Woods:2026p206201, Woods2024}, where the two approaches were confirmed to give results in excellent agreement. 
In the $\{\ket{+z},\ket{-z}\}$ valley basis, the effective-mass Hamiltonian takes the form
\begin{equation}
    H_{\text{em}}=
    \begin{pmatrix}
        H_0 & H_v \\
        H_v^\dagger & H_0
    \end{pmatrix},
    \label{eq:EM_hamiltonian}
\end{equation}
where $H_0$ is the intravalley Hamiltonian and $H_v$ 
contains the intervalley couplings. The intravalley Hamiltonian is given by
\begin{equation}
    H_0 =
    -\frac{\hbar^2}{2m_l}\partial_z^2
    +V_{\mathrm{qw}}(z),
    \label{eq:intravalley_hamiltonian}
\end{equation}
where $m_l=0.92m_e$ is the longitudinal effective mass. The 
quantum-well confinement potential includes a uniform vertical electric field $F_z$, and is given by
\begin{equation}
    V_{\mathrm{qw}}(z)=E_{\mathrm{Ge}}n_{\mathrm{Ge}}(z)+eF_z z.
    \label{eq:qw_profile}
\end{equation}
Here we adopt the value $E_{\mathrm{Ge}}=0.6~\mathrm{eV}$, as consistent with \cite{Stern92}. 
However, other sources suggest a slightly smaller value of $E_\text{Ge}\approx 0.5$~eV \cite{Schaffler1997}, which would allow the wave function to penetrate further into the quantum-well barriers.
The Ge concentration profile is 
modeled using smooth sigmoidal interfaces:
\begin{equation}
\begin{aligned}
n_{\mathrm{Ge}}(z)={}&\bar{n}_{\mathrm{Ge}}
\left[
2-\frac{1}{1+\exp\left(-\dfrac{z+L/2}{w/4}\right)}
\right.\\
&\left.
-\frac{1}{1+\exp\left(\dfrac{z-L/2}{w/4}\right)}
\right] ,
\end{aligned}
\label{eq:ge_profile}
\end{equation}
where $L$ is the quantum-well length, $w$ characterizes the interface 
width, and $\bar{n}_{\mathrm{Ge}}=0.3$ is the nominal Ge concentration in the 
barriers. Both $L$ and $w$ are varied throughout this work.

The intervalley Hamiltonian comprises a spin-conserving term and a spin-orbit contribution:
\begin{equation}
\begin{aligned}
H_{v} ={}&
V_{\mathrm{qw}}(z)e^{-2ik_0z} \\
&-i\varepsilon_{xy}e^{2ik_1z}
\left[
C_1+C_2\left(
2ik_1\partial_z+\partial_z^2
\right)
\right] \\
&+\beta_0e^{2ik_1z}
(\hat{k}_x\sigma_x - \hat{k}_y\sigma_y).
\end{aligned}
\label{eq:valley_coupling_operator}
\end{equation}
We adopt the values $C_1=1.73~\mathrm{eV}$ and $C_2=0.032a_0^2C_1$, derived in \cite{Woods2024}
\footnote{The value of $C_2$ reported here corrects a factor-of-2 discrepancy in Ref.~\cite{Woods2024}.
However, we note that the $C_2$ term in Eq.~(\ref{eq:valley_coupling_operator}) has almost no effect on the valley splitting, compared to the $C_1$ term.}.
The first term in Eq.~\eqref{eq:valley_coupling_operator} describes conventional potential-induced intervalley coupling with momentum transfer $2k_0$,
whereas the second describes the $2k_1$ intervalley coupling enabled by shear strain. The final term in Eq.~(\ref{eq:valley_coupling_operator}) describes the intervalley Dresselhaus spin-orbit interaction, where $\beta_0=8.2~\mathrm{meV\,nm}$ is the bare Dresselhaus coupling coefficient for Si obtained from a multiband tight-binding theory \cite{Woods:2026p206201}.

We numerically solve Eq.~\eqref{eq:intravalley_hamiltonian} to obtain
the envelope functions $\varphi_\nu(z)$ and energies $\varepsilon_\nu$ for each subband $\nu$, with emphasis on the dominant ground subband, $\nu=0$.
The spin-conserving intervalley matrix element for $\nu=0$ (also known as the valley coupling) is given by $\Delta = \mel{\varphi_0}{H_v}{\varphi_0}$, and the corresponding valley splitting is given by $E_v = 2|\Delta|$. 
We note that the effective-mass Hamiltonian inherits lateral spatial variations from $V_\text{qw}$ and $\varepsilon_{xy}$.
The ground-state intrasubband Dresselhaus coefficient is obtained by projecting the third term in Eq.~(\ref{eq:valley_coupling_operator}) onto the lowest subband: 
\begin{equation}
\beta_{00} = \beta_0 \mel{\varphi_0}{e^{i2k_1z}}{\varphi_0}. \label{eq:beta00}
\end{equation}
The magnitude $|\beta_{00}|$ therefore quantifies the effective intrasubband spin-orbit coupling relevant for electrically driven spin rotations (EDSR) \cite{soomro2026preprint}.

Following Ref.~\cite{Woods:2026p206201}, we note that spin-orbit coupling also renormalizes the $g$-factors of the two valley eigenstates. 
Except for certain anomalous cases, the renormalized $g$-factors are given by
\begin{equation}
    g_\pm \simeq g_0 \mp |g_\tau| \cos(\phi_v-\phi_g) \sin(2\theta_B) ,
    \label{eq:g_factor}
\end{equation}
for the ground (excited) valley states, where $\phi_v=\arg(\Delta)$ is the valley phase, $\phi_g=\arg(g_\tau)$ is a $g$-factor phase, and $\theta_B$ is
the in-plane angle of the magnetic field, relative to the $[100]$ crystallographic axis. The intervalley $g$-factor renormalization parameter is given by
\begin{equation}
    g_\tau = \frac{2e\hbar}{\mu_Bm_l}
    \sum_{\nu>0}
    \frac{\left(ik_z^{0\nu}\right)\beta_{\nu0}}
    {\varepsilon_\nu-\varepsilon_0+\hbar\omega_t},
    \label{eq:g_tau}
\end{equation}
where $\hbar\omega_t$ is the in-plane orbital confinement energy, which we take to be $\hbar\omega_t=$1~meV throughout this work, and we define $\beta_{\mu,\nu} = \beta_0\bra{\mu}e^{i2k_1 z}\ket{\nu}$ and $k_z^{\mu, \nu} = \bra{\mu}\hat k_z\ket{\nu}$. 
Based on these definitions, we note that $\beta_{00}$ describes the intrasubband spin-orbit coupling, while $g_\tau$ arises from virtual transitions involving higher subbands and consequently depends on the intersubband coefficients $\beta_{\nu0}$.

\bibliography{bibliography}

@article{
deLeon:2021peabb2823,
author = {Nathalie P. de Leon  and Kohei M. Itoh  and Dohun Kim  and Karan K. Mehta  and Tracy E. Northup  and Hanhee Paik  and B. S. Palmer  and N. Samarth  and Sorawis Sangtawesin  and D. W. Steuerman },
title = {Materials challenges and opportunities for quantum computing hardware},
journal = {Science},
volume = {372},
number = {6539},
pages = {eabb2823},
year = {2021},
doi = {10.1126/science.abb2823},
URL = {https://www.science.org/doi/abs/10.1126/science.abb2823}
}

@article{Schaffler1997,
doi = {10.1088/0268-1242/12/12/001},
url = {https://dx.doi.org/10.1088/0268-1242/12/12/001},
year = {1997},
month = {dec},
publisher = {},
volume = {12},
number = {12},
pages = {1515},
author = {Friedrich Schäffler},
title = {High-mobility {Si} and {Ge} structures},
journal = {Semiconductor Science and Technology}}

@article{Boykin:2004p115,
 affiliation = {Univ Alabama, Dept Elect {\&} Comp Engn, Huntsville, AL 35899 USA},
 author = {Boykin, T. B. and Klimeck, G. and Eriksson, M. A. and Friesen, M. and Coppersmith, S. N. and von Allmen, P. and Oyafuso, F. and Lee, S.},
 doi = {10.1063/1.1637718},
 journal = {Appl. Phys. Lett.},
 pages = {115--117},
 title = {Valley splitting in strained silicon quantum wells},
 volume = {84},
 year = {2004}
}

@article{Dyck:2017p1700622,
 author = {Dyck, O. and Leonard, D. N. and Edge, L. F. and Jackson, C. A. and Pritchett, E. J. and Deelman, P. W. and Poplawsky, J. D.},
 journal = {Adv. Mater. Interfaces},
 doi = {10.1002/admi.201700622},
 pages = {1700622},
 title = {Accurate Quantification of {Si/SiGe} Interface Profiles via Atom Probe Tomography},
 volume = {4},
 year = {2017}
}

@book{bradley2013,
  author = {Peter Bradley and Ray Radebaugh},
  title = {Properties of Selected Materials at Cryogenic Temperatures},
  year = {2013},
  publisher = {CRC Press, Boca Raton, FL},
  url = {https://tsapps.nist.gov/publication/get_pdf.cfm?pub_id=913059}
}

@article{Burkard:2023p025003,
  title = {Semiconductor spin qubits},
  author = {Burkard, Guido and Ladd, Thaddeus D. and Pan, Andrew and Nichol, John M. and Petta, Jason R.},
  journal = {Rev. Mod. Phys.},
  volume = {95},
  issue = {2},
  pages = {025003},
  numpages = {58},
  year = {2023},
  month = {Jun},
  publisher = {American Physical Society},
  doi = {10.1103/RevModPhys.95.025003},
  url = {https://link.aps.org/doi/10.1103/RevModPhys.95.025003}
}

@article{Burke1962ThermalC,
  author  = {Burke, M.},
  title   = {Thermal expansion of ceramic materials at $-200^\circ$ to $0^\circ${C}},
  journal = {Journal of the American Ceramic Society},
  year    = {1962},
  volume  = {45},
  pages   = {305--306}
}

@book{Casella2002,
  author    = {George Casella and Roger L. Berger},
  title     = {Statistical Inference},
  edition   = {2},
  year      = {2002},
  publisher = {Duxbury Press},
  address   = {Pacific Grove, CA}
}

@article{Cai:2023p386,
	Author = {Cai, Xinxin and Connors, Elliot J. and Edge, Lisa F. and Nichol, John M.},
	Da = {2023/03/01},
	Doi = {10.1038/s41567-022-01870-y},
	Id = {Cai2023},
	Isbn = {1745-2481},
	Journal = {Nature Physics},
	Number = {3},
	Pages = {386--393},
	Title = {Coherent spin--valley oscillations in silicon},
	Ty = {JOUR},
	Url = {https://doi.org/10.1038/s41567-022-01870-y},
	Volume = {19},
	Year = {2023}}

@misc{rahlff2026preprint,
      title={Sharp periodic {Ge} concentration modulations beyond the conduction band valley wavevector $k_0$ in nuclear spin-free {Si} quantum wells}, 
      author={Ivo Rahlff and Carsten Richter and Martin Schmidbauer and Maximilian Oezkent and Thilo Remmele and Michael Hanke and Lars R. Schreiber and Denny Dütz and Sammy Umezawa and Martin Albrecht and Yann-Michel Niquet and Tancredi Salamone and Biel Martinez Diaz and Thomas Schroeder and Jens Martin and Kevin-P. Gradwohl},
      year={2026},
      eprint={2605.31358},
      archivePrefix={arXiv},
      url={https://arxiv.org/abs/2605.31358}, 
}

@article{Koch:2025p59,
  author = {Koch, Thomas and Godfrin, Clement and Adam, Viktor and Ferrero, Julian and Schroller, Daniel and Glaeser, Noah and Kubicek, Stefan and Li, Ruoyu and Loo, Roger and Massar, Shana and Simion, George and Wan, Danny and De Greve, Kristiaan and Wernsdorfer, Wolfgang},
  title = {Industrial 300 mm wafer processed spin qubits in natural silicon/silicon-germanium},
  journal = {npj Quantum Information},
  year = {2025},
  month = {apr},
  volume = {11},
  number = {1},
  pages = {59},
  doi = {10.1038/s41534-025-01016-x},
  issn = {2056-6387},
}

@article{Chen:2021p044033,
    title = {Detuning Axis Pulsed Spectroscopy of Valley-Orbital States in $\mathrm{Si}$/$\mathrm{Si}$-$\mathrm{Ge}$ Quantum Dots},
    author = {Chen, Edward H. and Raach, Kate and Pan, Andrew and Kiselev, Andrey A. and Acuna, Edwin and Blumoff, Jacob Z. and Brecht, Teresa and Choi, Maxwell D. and Ha, Wonill and Hulbert, Daniel R. and Jura, Michael P. and Keating, Tyler E. and Noah, Ramsey and Sun, Bo and Thomas, Bryan J. and Borselli, Matthew G. and Jackson, C.A.C. and Rakher, Matthew T. and Ross, Richard S.},
    journal = {Phys. Rev. Applied},
    volume = {15},
    issue = {4},
    pages = {044033},
    numpages = {13},
    year = {2021},
    month = {Apr},
    publisher = {American Physical Society},
    doi = {10.1103/PhysRevApplied.15.044033},
    url = {https://link.aps.org/doi/10.1103/PhysRevApplied.15.044033}
}

@article{Acuna:2024p044057,
  title = {Coherent control of a triangular exchange-only spin qubit},
  author = {Acuna, Edwin and Broz, Joseph D. and Shyamsundar, Kaushal and Mei, Antonio B. and Feeney, Colin P. and Smetanka, Valerie and Davis, Tiffany and Lee, Kangmu and Choi, Maxwell D. and Boyd, Brydon and Suh, June and Ha, Wonill and Jennings, Cameron and Pan, Andrew S. and Sanchez, Daniel S. and Reed, Matthew D. and Petta, Jason R.},
  journal = {Phys. Rev. Appl.},
  volume = {22},
  issue = {4},
  pages = {044057},
  numpages = {6},
  year = {2024},
  month = {Oct},
  publisher = {American Physical Society},
  doi = {10.1103/PhysRevApplied.22.044057},
  url = {https://link.aps.org/doi/10.1103/PhysRevApplied.22.044057}
}

@misc{Comsol, 
title = {{COMSOL} {M}ultiphysics\textsuperscript{\textregistered} v.6.4. \url{www.comsol.com}. {COMSOL AB}, {Stockholm}, {Sweden}},
year={2025}
}

@book{Corruccini1961ThermalLiterature,
  author    = {Corruccini, Robert J. and Gniewek, John J.},
  title     = {Thermal Expansion of Technical Solids at Low Temperatures:
               A Compilation From the Literature},
  series    = {NBS Monograph},
  number    = {29},
  publisher = {National Bureau of Standards},
  address   = {Washington, D.C.},
  year      = {1961},
  url       = {https://digital.library.unt.edu/ark:/67531/metadc70437/}
}

@article{Neyens2024p80,
  author = {Neyens, Samuel and Zietz, Otto K. and Watson, Thomas F. and Luthi, Florian and Nethwewala, Aditi and George, Hubert C. and Henry, Eric and Islam, Mohammad and Wagner, Andrew J. and Borjans, Felix and Connors, Elliot J. and Corrigan, J. and Curry, Matthew J. and Keith, Daniel and Kotlyar, Roza and Lampert, Lester F. and Mądzik, Mateusz T. and Millard, Kent and Mohiyaddin, Fahd A. and Pellerano, Stefano and Pillarisetty, Ravi and Ramsey, Mick and Savytskyy, Rostyslav and Schaal, Simon and Zheng, Guoji and Ziegler, Joshua and Bishop, Nathaniel C. and Bojarski, Stephanie and Roberts, Jeanette and Clarke, James S.},
  title = {Probing single electrons across 300-mm spin qubit wafers},
  journal = {Nature},
  year = {2024},
  month = {may},
  volume = {629},
  number = {8010},
  pages = {80--85},
  doi = {10.1038/s41586-024-07275-6},
  issn = {1476-4687},
}

@article{Feng:2022p085304,
  title = {Enhanced valley splitting in {Si} layers with oscillatory {Ge} concentration},
  author = {Feng, Yi and Joynt, Robert},
  journal = {Phys. Rev. B},
  volume = {106},
  issue = {8},
  pages = {085304},
  numpages = {11},
  year = {2022},
  month = {Aug},
  publisher = {American Physical Society},
  doi = {10.1103/PhysRevB.106.085304},
  url = {https://link.aps.org/doi/10.1103/PhysRevB.106.085304}
}

@article{Ferdous:2018p26,
    Author = {Ferdous, Rifat and Kawakami, Erika and Scarlino, Pasquale and Nowak, Micha{\l}P. and Ward, D. R. and Savage, D. E. and Lagally, M. G. and Coppersmith, S. N. and Friesen, Mark and Eriksson, Mark A. and Vandersypen, Lieven M. K. and Rahman, Rajib},
    Da = {2018/06/05},
    Doi = {10.1038/s41534-018-0075-1},
    Id = {Ferdous2018},
    Isbn = {2056-6387},
    Journal = {npj Quantum Information},
    Number = {1},
    Pages = {26},
    Title = {Valley dependent anisotropic spin splitting in silicon quantum dots},
    Ty = {JOUR},
    Url = {https://doi.org/10.1038/s41534-018-0075-1},
    Volume = {4},
    Year = {2018}
}

@article{Fischetti1996BandAlloys,
    title = {{Band structure, deformation potentials, and carrier mobility in strained Si, Ge, and SiGe alloys}},
    year = {1996},
    journal = {Journal of Applied Physics},
    author = {Fischetti, M V and Laux, S E},
    number = {4},
    month = {4},
    pages = {2234--2252},
    volume = {80},
    doi = {10.1063/1.363052},
    issn = {0021-8979}
}

@article{Friesen:2007p115318,
 author = {M Friesen and S Chutia and C Tahan and S N Coppersmith},
 journal = {Phys. Rev. B},
 pages = {115318},
 title = {Valley splitting theory of {S}i{G}e/{S}i/{S}i{G}e quantum wells},
 volume = {75},
 year = {2007}
}

@article{Ha2025PRXQuantum,
  title     = {Two-Dimensional {Si} Spin Qubit Arrays with Multilevel Interconnects},
  author    = {Ha, Sieu D. and Acuna, Edwin and Raach, Kate and Bloom, Zachery T. and Brecht, Teresa L. and Chappell, James M. and Choi, Maxwell D. and Christensen, Justin E. and Counts, Ian T. and Daprano, Dominic and Dodson, J. P. and Eng, Kevin and Fialkow, David J. and Garcia, Christina A. C. and Ha, Wonill and Harris, Thomas R. B. and Holman, Nathan and Khalaf, Isaac and Matten, Justine W. and Peterson, Christi A. and Plesha, Clifford E. and Ruiz, Matthew J. and Smith, Aaron and Thomas, Bryan J. and Whiteley, Samuel J. and Ladd, Thaddeus D. and Jura, Michael P. and Rakher, Matthew T. and Borselli, Matthew G.},
  journal   = {PRX Quantum},
  volume    = {6},
  issue     = {3},
  pages     = {030327},
  year      = {2025},
  month     = {Aug},
  publisher = {American Physical Society},
  doi       = {10.1103/sgn1-1t2d},
  url       = {https://link.aps.org/doi/10.1103/sgn1-1t2d}
}

@article{Jones2019,
  title = {Spin-Blockade Spectroscopy of $\mathrm{Si}$/$\mathrm{Si}$-$\mathrm{Ge}$ Quantum Dots},
  author = {Jones, A.M. and Pritchett, E.J. and Chen, E.H. and Keating, T.E. and Andrews, R.W. and Blumoff, J.Z. and De Lorenzo, L.A. and Eng, K. and Ha, S.D. and Kiselev, A.A. and Meenehan, S.M. and Merkel, S.T. and Wright, J.A. and Edge, L.F. and Ross, R.S. and Rakher, M.T. and Borselli, M.G. and Hunter, A.},
  journal = {Phys. Rev. Appl.},
  volume = {12},
  issue = {1},
  pages = {014026},
  numpages = {12},
  year = {2019},
  month = {Jul},
  publisher = {American Physical Society},
  doi = {10.1103/PhysRevApplied.12.014026},
  url = {https://link.aps.org/doi/10.1103/PhysRevApplied.12.014026}
}

@article{Kim:2000p577,
  author  = {Kim, Young-Soo and Kim, Jun Ki and Park, Jin Won and Kim, Do-Heyoung and Lee, Wan-Jin and Cho, Dong-Lyun and Song, Yo Soon},
  title   = {Properties of Chemically Vapor Deposited Blanket Tungsten Films
on TiN Glue Layers Prepared by Chemical Vapor Deposition},
  journal = {Metals and Materials},
  year    = {2000},
  volume  = {6},
  number  = {6},
  pages   = {577--581},
  doi     = {https://doi.org/10.1007/BF03028102}
}

@article{LEE2026166754,
title = {Effect of {W} nucleation process on the adhesion of {W/TiN/SiO}$_2$ structures for {BEOL} interconnects},
journal = {Applied Surface Science},
volume = {736},
pages = {166754},
year = {2026},
issn = {0169-4332},
doi = {https://doi.org/10.1016/j.apsusc.2026.166754},
url = {https://www.sciencedirect.com/science/article/pii/S016943322600958X},
author = {Sun Woo Lee and Dong Jun Kim and Jinho Jeon and Taek-Soo Kim}
}

@book{Levinshtein2000HandbookInSb,
  title={Properties of Advanced Semiconductor Materials: GaN, AIN, InN, BN, SiC, SiGe},
  author={Levinshtein, M.E. and Rumyantsev, S.L. and Shur, M.S.},
  isbn={9780471358275},
  lccn={00042262},
  series={A Wiley-Interscience publication},
  url={https://books.google.com/books?id=u26CpULkD_wC},
  year={2001},
  publisher={Wiley}
}

@article{Losert:2023p125405,
  title = {Practical strategies for enhancing the valley splitting in {S}i/{S}i{G}e quantum wells},
  author = {Losert, Merritt P. and Eriksson, M. A. and Joynt, Robert and Rahman, Rajib and Scappucci, Giordano and Coppersmith, Susan N. and Friesen, Mark},
  journal = {Phys. Rev. B},
  volume = {108},
  issue = {12},
  pages = {125405},
  numpages = {31},
  year = {2023},
  month = {Sep},
  publisher = {American Physical Society},
  doi = {10.1103/PhysRevB.108.125405},
  url = {https://link.aps.org/doi/10.1103/PhysRevB.108.125405}
}

@article{McJunkin:2021p085406,
  title = {Valley splittings in {S}i/{S}i{G}e quantum dots with a germanium spike in the silicon well},
  author = {McJunkin, Thomas and MacQuarrie, E. R. and Tom, Leah and Neyens, S. F. and Dodson, J. P. and Thorgrimsson, Brandur and Corrigan, J. and Ercan, H. Ekmel and Savage, D. E. and Lagally, M. G. and Joynt, Robert and Coppersmith, S. N. and Friesen, Mark and Eriksson, M. A.},
  journal = {Phys. Rev. B},
  volume = {104},
  issue = {8},
  pages = {085406},
  numpages = {11},
  year = {2021},
  month = {Aug},
  publisher = {American Physical Society},
  doi = {10.1103/PhysRevB.104.085406},
  url = {https://link.aps.org/doi/10.1103/PhysRevB.104.085406}
}

@article{McJunkin:2022p7777,
      title={Si{G}e quantum wells with oscillating {Ge} concentrations for quantum dot qubits}, 
      author={Thomas McJunkin and Benjamin Harpt and Yi Feng and Merritt Losert and Rajib Rahman and J. P. Dodson and M. A. Wolfe and D. E. Savage and M. G. Lagally and S. N. Coppersmith and Mark Friesen and Robert Joynt and M. A. Eriksson},
  journal = {Nature Commun.},
  volume = {13},
  pages = {7777},
  numpages = {7},
  year = {2022},
  month = {Dec},
  doi = {10.1038/s41467-022-35510-z},
}

@article{McSkimin1953,
   author = {H. J. McSkimin},
   doi = {10.1121/1.1917687},
   issn = {0001-4966},
   issue = {4},
   journal = {The Journal of the Acoustical Society of America},
   month = {7},
   pages = {826-826},
   title = {Measurement of Elastic Constants at Low Temperatures by Means of High Frequency Ultrasonic Waves},
   volume = {25},
   year = {1953}
}

@article{Nestoklon:2006p235334,
 affiliation = {Russian Acad Sci, AF Ioffe Physicotech Inst, St Petersburg 194021, Russia},
 author = {M O Nestoklon and L E Golub and E L Ivchenko},
 doi = {10.1103/PhysRevB.73.235334},
 journal = {Phys Rev B},
 month = {Jan},
 number = {23},
 pages = {235334},
 title = {Spin and valley-orbit splittings in {SiGe/Si} heterostructures},
 volume = {73},
 year = {2006}
}

@article{Prada:2011p013009,
doi = {10.1088/1367-2630/13/1/013009},
url = {https://dx.doi.org/10.1088/1367-2630/13/1/013009},
year = {2011},
month = {jan},
publisher = {},
volume = {13},
number = {1},
pages = {013009},
author = {M Prada and G Klimeck and R Joynt},
title = {Spin-orbit splittings in {Si/SiGe} quantum wells: from ideal {Si} membranes to realistic heterostructures},
journal = {New Journal of Physics},
}

@article{Schaffler:1991p2039,
    author = {F. Sch\"{a}ffler and H.-J. Herzog and H. Jorke and E. Kasper},
    title = {Influence of thermal annealing on the electron mobility in modulation doped {S}i/{S}i{G}e heterostructures},
    journal = {J. Vac. Sci. Technol. B},
    volume = {9},
    pages = {2039},
    year = {1991}
}

@book{Simon1994,
title = {Cryogenic Properties of Inorganic Insulation Materials for ITER Magnets: A Review},
author = {Simon, N. J.},
publisher = {National Institute of Standards and Technology, Boulder, CO},
doi = {10.2172/761710},
url = {https://www.osti.gov/biblio/761710}, 
place = {United States},
year = {1994},
month = {12}
}

@misc{Emily2026preprint,
      title={Enhancing valley splitting in {Si/SiGe Wiggle Well} quantum dots through integrated stressor-gate architectures}, 
      author={E. S. Joseph and Talise Oh and D. E. Savage and D. S. Stone and DeAnna Campbell and Benjamin D. Woods and Mark Friesen and M. A. Eriksson},
      year={2026},
      eprint={2611.12850},
      archivePrefix={arXiv},
      url={https://arxiv.org/abs/2611.12850}, 
}

@article{Thayil:2025p115303,
  title = {Theory of valley splitting in {Si/SiGe} spin qubits: Interplay of strain, resonances, and random alloy disorder},
  author = {Thayil, Abel and Ermoneit, Lasse and Kantner, Markus},
  journal = {Phys. Rev. B},
  volume = {112},
  issue = {11},
  pages = {115303},
  numpages = {18},
  year = {2025},
  month = {Sep},
  publisher = {American Physical Society},
  doi = {10.1103/4sdz-f9cr},
  url = {https://link.aps.org/doi/10.1103/4sdz-f9cr}
}

@article{Nemeth:2026p010336,
  title = {Omnidirectional Shuttling to Avoid Valley Excitations in $\mathrm{Si}$/$\mathrm{Si}\mathrm{Ge}$ Quantum Wells},
  author = {N\'emeth, R\'obert and Bandaru, Vatsal K. and Alves, Pedro and Brann, Emma and Eskandari, Owen M. and Soomro, Hudaiba and Vivrekar, Avani and Eriksson, M.A. and Losert, Merritt P. and Friesen, Mark},
  journal = {PRX Quantum},
  volume = {7},
  issue = {1},
  pages = {010336},
  numpages = {17},
  year = {2026},
  month = {Feb},
  publisher = {American Physical Society},
  doi = {10.1103/615j-xjyh},
  url = {https://link.aps.org/doi/10.1103/615j-xjyh}
}

@misc{king2026preprint,
      title={Complete measurement of tunnel- and valley-coupling parameters in a silicon double quantum dot}, 
      author={Daniel J. King and Minyoung Kim and J. Reily and Jonathan C. Marcks and Mark Friesen and Benjamin D. Woods and M. A. Eriksson},
      year={2026},
      eprint={2607.09638},
      archivePrefix={arXiv},
      url={https://arxiv.org/abs/2607.09638}, 
}

@misc{woods2025preprint,
      title={Statistical characterization of valley coupling in {Si/SiGe} quantum dots via $g$-factor measurements near a valley vortex}, 
      author={Benjamin D. Woods and Merritt P. Losert and Nasir R. Elston and Hudaiba Soomro and M. A. Eriksson and S. N. Coppersmith and Robert Joynt and Mark Friesen},
      year={2026},
      eprint={2507.05160},
      archivePrefix={arXiv},
      url={https://arxiv.org/abs/2507.05160}, 
}

@article{Thorbeck:2015p087107,
    author = {Ted Thorbeck and Neil M. Zimmerman},
    title = {Formation of strain-induced quantum dots in gated semiconductor nanostructures},
    journal = {AIP Advances},
    volume = {5},
    pages = {087107},
    year = {2015}
}

@article{Volmer:2024p61,
	Author = {Volmer, Mats and Struck, Tom and Sala, Arnau and Chen, Bingjie and Oberl{\"a}nder, Max and Offermann, Tobias and Xue, Ran and Visser, Lino and Tu, Jhih-Sian and Trellenkamp, Stefan and Cywi{\'n}ski, {\L}ukasz and Bluhm, Hendrik and Schreiber, Lars R.},
	Da = {2024/06/19},
	Doi = {10.1038/s41534-024-00852-7},
	Id = {Volmer2024},
	Isbn = {2056-6387},
	Journal = {npj Quantum Information},
	Number = {1},
	Pages = {61},
	Title = {Mapping of valley splitting by conveyor-mode spin-coherent electron shuttling},
	Ty = {JOUR},
	Volume = {10},
	Year = {2024}
}

@misc{volmer2026preprint,
      title={Mapping $g$-factors and complex intervalley coupling in {Si/SiGe} by conveyor-mode shuttling}, 
      author={Mats Volmer and Tom Struck and Arnau Sala and Jhih-Sian Tu and Stefan Trellenkamp and Davide Degli Esposti and Giordano Scappucci and Hendrik Bluhm and Łukasz Cywiński and Lars R. Schreiber},
      year={2026},
      eprint={2603.01844},
      archivePrefix={arXiv},
      url={https://arxiv.org/abs/2603.01844}, 
}

@article{Woods:2023p035418,
  title = {Spin-orbit enhancement in {S}i/{S}i{G}e heterostructures with oscillating {G}e concentration},
  author = {Woods, Benjamin D. and Eriksson, M. A. and Joynt, Robert and Friesen, Mark},
  journal = {Phys. Rev. B},
  volume = {107},
  issue = {3},
  pages = {035418},
  numpages = {17},
  year = {2023},
  month = {Jan},
  publisher = {American Physical Society},
  doi = {10.1103/PhysRevB.107.035418},
  url = {https://link.aps.org/doi/10.1103/PhysRevB.107.035418}
}

@misc{soomro2026preprint,
      title={High-fidelity {EDSR} in {Si/SiGe} {W}iggle {W}ells}, 
      author={Hudaiba Soomro and Minyoung Kim and Avani Vivrekar and M. A. Eriksson and Benjamin D. Woods and Mark Friesen},
      year={2026},
      eprint={2605.24790},
      archivePrefix={arXiv},
      url={https://arxiv.org/abs/2605.24790}, 
}

@article{Woods2024,
  author  = {Woods, Benjamin D. and Soomro, Hudaiba and Joseph, E. S. and Frink, Collin C. D. and Joynt, Robert and Eriksson, M. A. and Friesen, Mark},
  title   = {Coupling conduction-band valleys in {SiGe} heterostructures via shear strain and {Ge} concentration oscillations},
  journal = {npj Quantum Information},
  volume  = {10},
  pages   = {54},
  year    = {2024},
  doi     = {10.1038/s41534-024-00853-6}
}

@misc{frink2025preprint,
      title={Reducing strain fluctuations in quantum dot devices by gate-layer stacking}, 
      author={Collin C. D. Frink and Talise Oh and E. S. Joseph and Merritt P. Losert and E. R. MacQuarrie and Benjamin D. Woods and M. A. Eriksson and Mark Friesen},
      year={2025},
      eprint={2312.09235},
      archivePrefix={arXiv},
      url={https://arxiv.org/abs/2312.09235}, 
}

@article{Woods:2026p206201,
  title = {$g$-Factor Theory of {Si/SiGe} Quantum Dots: Spin-Valley and Giant Renormalization Effects},
  author = {Woods, Benjamin D. and Losert, Merritt P. and Joynt, Robert and Friesen, Mark},
  journal = {Phys. Rev. Lett.},
  volume = {136},
  issue = {20},
  pages = {206201},
  numpages = {8},
  year = {2026},
  month = {May},
  publisher = {American Physical Society},
  doi = {10.1103/rd84-5g2x},
  url = {https://link.aps.org/doi/10.1103/rd84-5g2x}
}

@ARTICLE{Marcks:2025p11381,
  title    = "Valley splitting correlations across a silicon quantum well
              containing germanium",
  author   = "Marcks, Jonathan C and Eagen, Emily and Brann, Emma C and Losert,
              Merritt P and Oh, Talise and Reily, J and Wang, Christopher S and
              Keith, Daniel and Mohiyaddin, Fahd A and Luthi, Florian and
              Curry, Matthew J and Zhang, Jiefei and Heremans, F Joseph and
              Friesen, Mark and Eriksson, M A",
  journal  = "Nature Communications",
  volume   =  16,
  number   =  1,
  pages    = "11381",
  month    =  dec,
  year     =  2025
}

@article{Wuetz:2022p7730,
	Author = {Paquelet Wuetz, Brian and Losert, Merritt P. and Koelling, Sebastian and Stehouwer, Lucas E. A. and Zwerver, Anne-Marije J. and Philips, Stephan G. J. and M{\k a}dzik, Mateusz T. and Xue, Xiao and Zheng, Guoji and Lodari, Mario and Amitonov, Sergey V. and Samkharadze, Nodar and Sammak, Amir and Vandersypen, Lieven M. K. and Rahman, Rajib and Coppersmith, Susan N. and Moutanabbir, Oussama and Friesen, Mark and Scappucci, Giordano},
	Da = {2022/12/13},
	Doi = {10.1038/s41467-022-35458-0},
	Id = {Paquelet Wuetz2022},
	Isbn = {2041-1723},
	Journal = {Nature Communications},
	Number = {1},
	Pages = {7730},
	Title = {Atomic fluctuations lifting the energy degeneracy in {S}i/{S}i{G}e quantum dots},
	Ty = {JOUR},
	Volume = {13},
	Year = {2022}
}

@article{Ylivaara2022ThermomechanicalDeposition,
    title = {{Thermomechanical properties of aluminum oxide thin films made by atomic layer deposition}},
    year = {2022},
    journal = {Journal of Vacuum Science {\&} Technology A},
    author = {Ylivaara, Oili M E and Langner, Andreas and Ek, Satu and Malm, Jari and Julin, Jaakko and Laitinen, Mikko and Ali, Saima and Sintonen, Sakari and Lipsanen, Harri and Sajavaara, Timo and Puurunen, Riikka L},
    number = {6},
    month = {4},
    pages = {62414},
    volume = {40},
    url = {https://doi.org/10.1116/6.0002095},
    doi = {10.1116/6.0002095},
    issn = {0734-2101}
}

@article{Abraham:2026p1154,
  author = {Abraham, Michael and {Members of the HRL Quantum Team and Collaborators}},
  title = {A digitally controlled silicon quantum processing unit},
  journal = {Nature},
  year = {2026},
  month = {jul},
  volume = {655},
  number = {8125},
  pages = {1154--1159},
  doi = {10.1038/s41586-026-10754-7},
  issn = {1476-4687},
}

@article{Esposti2026,
    author = {Degli Esposti, Davide and Brann, Emma C. and Elsayed, Asser and Costa, Davide and Friesen, Mark and Scappucci, Giordano},
    title = {Confinement
Drives Valley Splitting above 4{K} in Buried
Silicon Quantum Wells},
    journal = {Nano Letters},
    volume = {26},
    number = {43},
    pages = {15131-15139},
    year = {2026},
    month = {11},
    issn = {1530-6984},
    doi = {10.1021/acs.nanolett.6c04172},
    url = {https://doi.org/10.1021/acs.nanolett.6c04172}
}

@article{Zajac:2018p439,
 author = {Zajac, D. M. and Sigillito, A. J. and Russ, M. and Borjans, F. and Taylor, J. M. and Burkard, G. and Petta, J. R.},
 journal = {Science},
 title = {Resonantly driven {CNOT} gate for electron spins},
 volume = {359},
 pages = {439},
 year = {2018}
}

@article{John:2025p10560,
  author = {John, Valentin and Yu, Cécile X. and van Straaten, Barnaby and Rodríguez-Mena, Esteban A. and Rodríguez, Mauricio and Oosterhout, Stefan D. and Stehouwer, Lucas E. A. and Scappucci, Giordano and Rimbach-Russ, Maximilian and Bosco, Stefano and Borsoi, Francesco and Niquet, Yann-Michel and Veldhorst, Menno},
  title = {Robust and localised control of a 10-spin qubit array in germanium},
  journal = {Nature Communications},
  year = {2025},
  month = {nov},
  volume = {16},
  number = {1},
  pages = {10560},
  doi = {10.1038/s41467-025-65577-3},
  issn = {2041-1723},
}

@article{Unseld:2023p084002,
    author = {Unseld, F. K. and Meyer, M. and Mądzik, M. T. and Borsoi, F. and de Snoo, S. L. and Amitonov, S. V. and Sammak, A. and Scappucci, G. and Veldhorst, M. and Vandersypen, L. M. K.},
    title = {A {2D} quantum dot array in planar 28{Si/SiGe}},
    journal = {Applied Physics Letters},
    volume = {123},
    number = {8},
    pages = {084002},
    year = {2023},
    month = {08},
    issn = {0003-6951},
    doi = {10.1063/5.0160847},
    url = {https://doi.org/10.1063/5.0160847}
}

@article{Ha:2025p030327,
  title = {Two-Dimensional $\mathrm{Si}$ Spin Qubit Arrays with Multilevel Interconnects},
  author = {Ha, Sieu D. and Acuna, Edwin and Raach, Kate and Bloom, Zachery T. and Brecht, Teresa L. and Chappell, James M. and Choi, Maxwell D. and Christensen, Justin E. and Counts, Ian T. and Daprano, Dominic and Dodson, J.P. and Eng, Kevin and Fialkow, David J. and Garcia, Christina A. C. and Ha, Wonill and Harris, Thomas R. B. and holman, nathan and Khalaf, Isaac and Matten, Justine W. and Peterson, Christi A. and Plesha, Clifford E. and Ruiz, Matthew J. and Smith, Aaron and Thomas, Bryan J. and Whiteley, Samuel J. and Ladd, Thaddeus D. and Jura, Michael P. and Rakher, Matthew T. and Borselli, Matthew G.},
  journal = {PRX Quantum},
  volume = {6},
  issue = {3},
  pages = {030327},
  numpages = {7},
  year = {2025},
  month = {Aug},
  publisher = {American Physical Society},
  doi = {10.1103/sgn1-1t2d},
  url = {https://link.aps.org/doi/10.1103/sgn1-1t2d}
}

@article{Nickl:2026p5878,
  author = {Nickl, Andreas and Dumoulin Stuyck, Nard and Steinacker, Paul and Cifuentes, Jesus D. and Serrano, Santiago and Feng, MengKe and Vahapoglu, Ensar and Hudson, Fay E. and Chan, Kok Wai and Kubicek, Stefan and Jussot, Julien and Canvel, Yann and Beyne, Sofie and Shimura, Yosuke and Loo, Roger and Godfrin, Clement and Raes, Bart and Baudot, Sylvain and Wan, Danny and Laucht, Arne and Yang, Chih-Hwan and Lim, Wee Han and Saraiva, Andre and Escott, Christopher C. and De Greve, Kristiaan and Dzurak, Andrew S. and Tanttu, Tuomo},
  title = {Eight-qubit operation of a 300 mm {SiMOS} foundry-fabricated device},
  journal = {Nature Communications},
  year = {2026},
  month = {jul},
  volume = {17},
  number = {1},
  pages = {5878},
  doi = {10.1038/s41467-026-74597-6},
  issn = {2041-1723},
}

@article{Zwerver:2022p184,
	Author = {Zwerver, A. M. J. and Kr{\"a}henmann, T. and Watson, T. F. and Lampert, L. and George, H. C. and Pillarisetty, R. and Bojarski, S. A. and Amin, P. and Amitonov, S. V. and Boter, J. M. and Caudillo, R. and Correas-Serrano, D. and Dehollain, J. P. and Droulers, G. and Henry, E. M. and Kotlyar, R. and Lodari, M. and L{\"u}thi, F. and Michalak, D. J. and Mueller, B. K. and Neyens, S. and Roberts, J. and Samkharadze, N. and Zheng, G. and Zietz, O. K. and Scappucci, G. and Veldhorst, M. and Vandersypen, L. M. K. and Clarke, J. S.},
	Da = {2022/03/01},
	Doi = {10.1038/s41928-022-00727-9},
	Id = {Zwerver2022},
	Isbn = {2520-1131},
	Journal = {Nature Electronics},
	Number = {3},
	Pages = {184--190},
	Title = {Qubits made by advanced semiconductor manufacturing},
	Ty = {JOUR},
	Volume = {5},
	Year = {2022}
}

@article{Stern92,
    author = {Stern, Frank and Laux, Steven E.},
    title = {Charge transfer and low-temperature electron mobility in a strained {Si} layer in relaxed {Si}$_{1-x}${Ge}$_x$},
    journal = {Applied Physics Letters},
    volume = {61},
    number = {9},
    pages = {1110-1112},
    year = {1992},
    month = {08},
    issn = {0003-6951},
    doi = {10.1063/1.107684},
    url = {https://doi.org/10.1063/1.107684}
}

\clearpage

\renewcommand\thefigure{S\arabic{figure}}
\renewcommand\theequation{S\arabic{equation}}
\renewcommand\thesection{S\arabic{section}}
\setcounter{figure}{0}
\setcounter{equation}{0}
\setcounter{section}{0}
\onecolumngrid
\vspace{0.5in}
\begin{center}
\textbf{\large Supplementary information for ``The importance of shear strain in narrow Si/SiGe quantum wells''}
\end{center}

These Supplementary Notes and Figures provide details of the calculations reported in the main text.

\section{Gate design simulations and gate parameter sweeps}
\label{sec:gate_design}

\begin{figure}[b]
	\includegraphics[width= 16 cm]{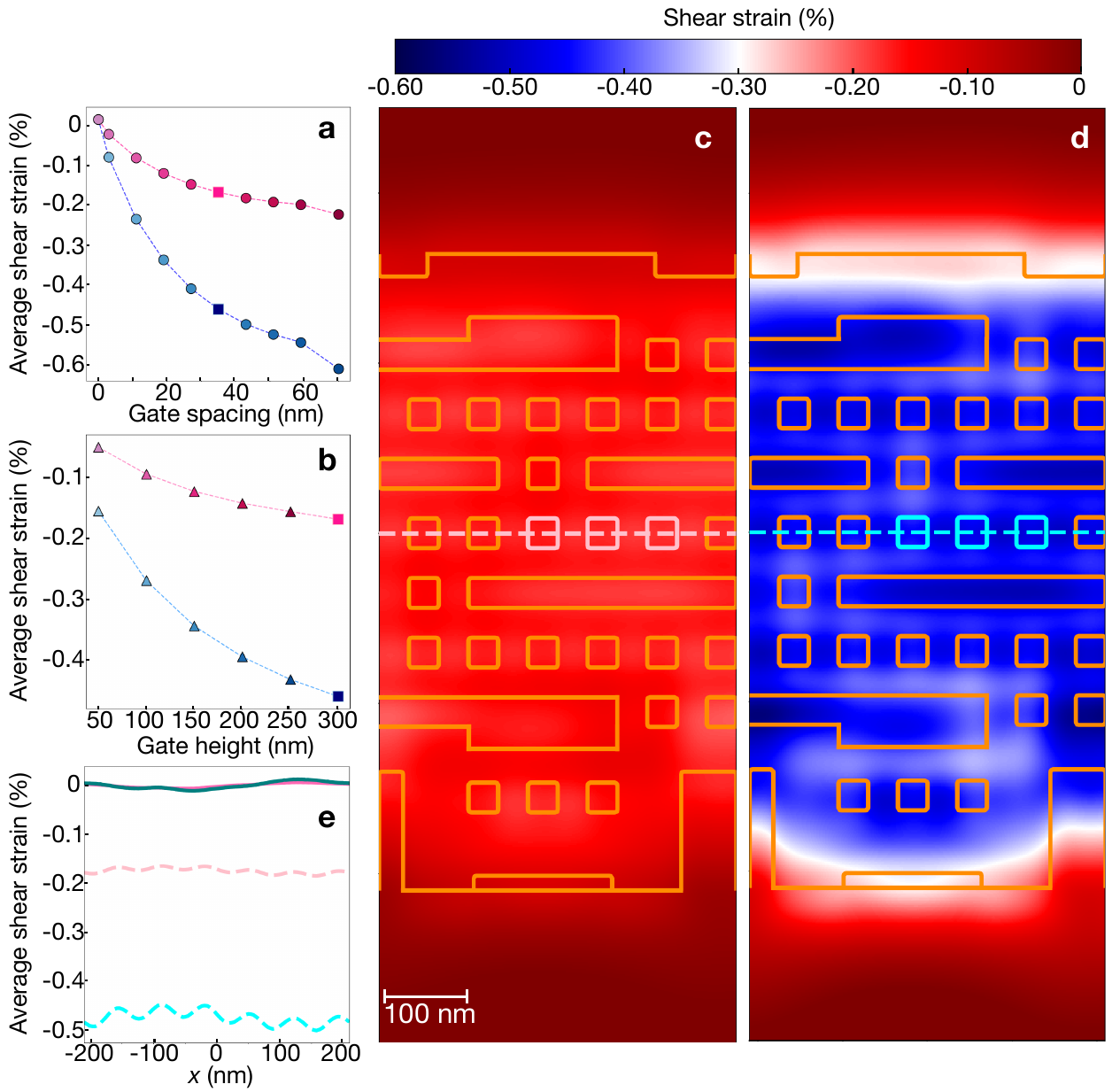}
	\centering
    \caption{\textbf{Shear strain data for variable gate spacing, gate height, and depositional stress.}
    \textbf{a},\textbf{b} Average shear strains are plotted for different gate geometries, as explained in Note~\ref{sec:gate_design}, as a function of \textbf{a} gate spacing, and \textbf{b} gate height. 
    Two  depositional stress values are used for the W gates: {$1.12 ~\mathrm{GPa}$} (pink) and {$4 ~\mathrm{GPa}$} (blue). 
    \textbf{a} Here we assume a fixed gate height of 300~nm. 
    Square markers indicate a gate spacing of 35~nm.  
    \textbf{b} Here we assume a fixed gate spacing of 35~nm. 
    Square markers indicate a gate height of 300~nm.  
    \textbf{c},\textbf{d} Shear-strain maps obtained for a geometry with a 35~nm gate spacing and a 300~nm gate height.
    The W depositional stress values in these two simulations are \textbf{c} $1.12 ~\mathrm{GPa}$, and \textbf{d} $4 ~\mathrm{GPa}$. 
    \textbf{e} The dashed pink and blue curves show linecuts obtained in \textbf{c} and \textbf{d}, under the pink and blue dashed lines.
    The dashed lines describe results for devices oriented along the [110] crystallographic axis.
    The solid lines correspond to the same linecuts, obtained for devices oriented along the [100] crystallographic axis, highlighting the fact that much larger shear-strain magnitudes are obtained for devices oriented along [110].}
	\label{fig:comsol_sims}
\end{figure}

The gate structure illustrated in Ref.~\cite{Abraham:2026p1154}, upon which our simulations are modeled, does not specify the full details of the gate geometry.
Indeed, Fig.~S24 of that work is only schematic, although the plunger gate pitch is specified as 70~nm.
Taking this as a starting point, we consider the gate spacing (i.e., the open area between the gates) to be a free parameter.
Moreover, the gates are formed of W and they are described as ``tall,'' taking the form of vertical vias.
We therefore also consider the total gate height to be a free parameter.
Although many other device parameters could be considered to be variable in this work, we limit our analysis to these two quantities.
In this Supplementary Note, we then compute the shear strain as a function of these parameters.
Since the depositional strain depends on many details of the fabrication procedure, we consider two values for the depositional stress of W reported in the literature: $1.12 ~\mathrm{GPa}$ \cite{Kim:2000p577} and $4 ~\mathrm{GPa}$ \cite{LEE2026166754}. 

\begin{figure}[b]
	\includegraphics[width= 17 cm]{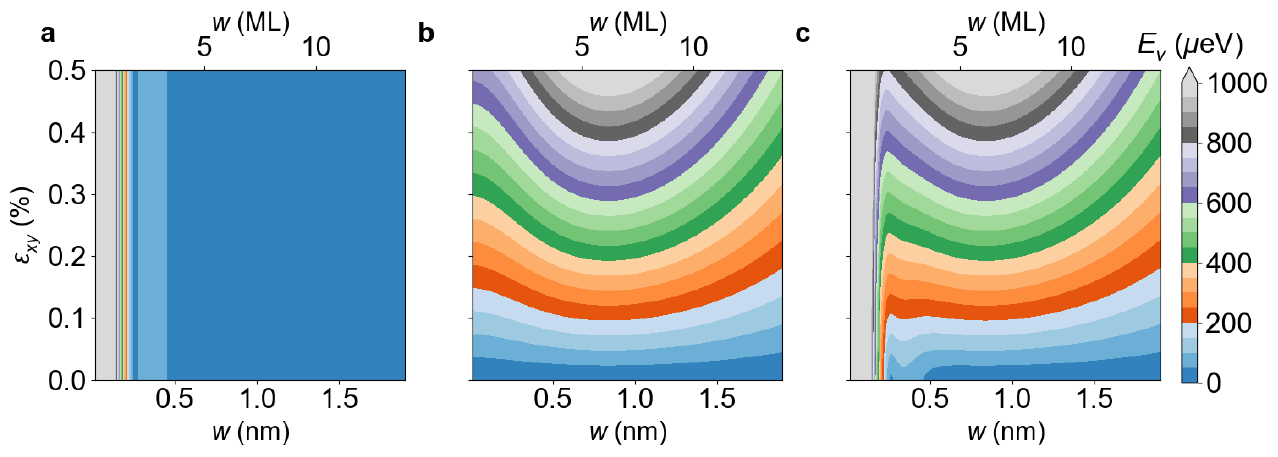}
	\centering
    \caption{
    \textbf{Individual contributions to the deterministic valley splitting from the $2k_0$ and $2k_1$ valley-coupling mechanisms.}
    This figure extends our understanding of Fig.~3a of the main text by separating the full results into their individual components.
    \textbf{a} The contribution to the valley splitting from the $2k_0$ coupling mechanism of Eq.~(\ref{eq:vs_2k_0}).
    This mechanism arises from ultrasharp features in the confinement potential, such as very small interface widths $w$.
    \textbf{b} The contribution to the valley splitting from the $2k_1$ coupling mechanism of Eq.~(\ref{eq:vs_2k_1}).
    This mechanism is proportional to the shear strain.
    \textbf{c} The total valley splitting $E_v$, which combines the $2k_0$ and $2k_1$ contributions.
    This panel is reproduced from Fig.~3a.}
	\label{fig:vs_k0_k1}
\end{figure}

We first explore a variable gate spacing, with a fixed value of 300~nm for the gate height.
In this case, we continuously vary the dimensions of the inner gates while retaining the overall gate pitch. 
However, we do not vary the topmost and bottommost gates in the gate schematic, which are dubbed ``outer field gates'' \cite{Abraham:2026p1154}.
Although the detailed shapes of these gates are not described in \cite{Abraham:2026p1154}, we assume they extend to the distant edges of the simulation cell, since ``field gates'' commonly refers to large-area gates used to deplete electrons, outside the active region \cite{Jones2019}.
Our approach to varying the plunger gate shapes is to change both gate dimensions simultaneously, as can be intuited by comparing the gate structure shown in Fig.~2b of the main text with Fig.~S24 of Ref.~\cite{Abraham:2026p1154}.
The results of these simulations are shown in Fig.~\ref{fig:comsol_sims}a, where the average shear strain is evaluated along the dashed lines shown in Figs.~\ref{fig:comsol_sims}c and d.
Here, the different colors represent the two depositional stress values assumed for W.
We observe a monotonic dependence of the shear strain on the gate spacing.
We note that the maximum gate spacing for this sweep is equal to the gate pitch (70~nm), and we emphasize that this limiting case is meaningful, in the sense that it corresponds to a geometry including outer field gates, but no plunger gates.
Since no optimal gate spacing stands out, within the sweep range, we simply adopt the middle value of 35~nm for the remaining simulations, including the simulations reported in the main text.
In Fig.~\ref{fig:comsol_sims}b, we also show results for gate-height sweeps.
In this case too, we find that no optimal value stands out. 
Hence, we settle on a physically reasonable \cite{Ha2025PRXQuantum} gate height of 300~nm for our remaining simulations.

Figures~\ref{fig:comsol_sims}c and d show shear-strain maps based on the gate parameters described in the previous paragraph.
Here, Fig.~\ref{fig:comsol_sims}c corresponds to the smaller depositional stress (1.12~GPa), while Fig.~\ref{fig:comsol_sims}d corresponds to the larger depositional stress (4~GPa).
Figure~\ref{fig:comsol_sims}c extends the results shown in Fig.~2b of the main text over a larger spatial range (and a different color scale), to demonstrate that the shear strain decays to zero as expected, far from the plunger gates.
In Fig.~\ref{fig:comsol_sims}e, we plot linecuts of the results obtained along the dashed lines in Figs.~\ref{fig:comsol_sims}c and d (shown as pink and blue dashed lines).
All the results discussed so far correspond to devices oriented along the [110] crystallographic direction.
It is important to note that the shear strain $\varepsilon_{xy}$ depends strongly on the device orientation. 
To highlight this point, in Fig.~\ref{fig:comsol_sims}e, we also report the same linecuts as before (solid lines), obtained along a line of plunger gates, but for a device oriented along [100].
The resulting shear strain values are very close to zero, demonstrating the importance of device orientation.

\section{Individual contributions from the \texorpdfstring{$2k_0$}{2k0} and \texorpdfstring{$2k_1$}{2k1} valley-coupling mechanisms}
\label{sec:vs_k0_k1}

\begin{figure}[b]
	\includegraphics[width= 18 cm]{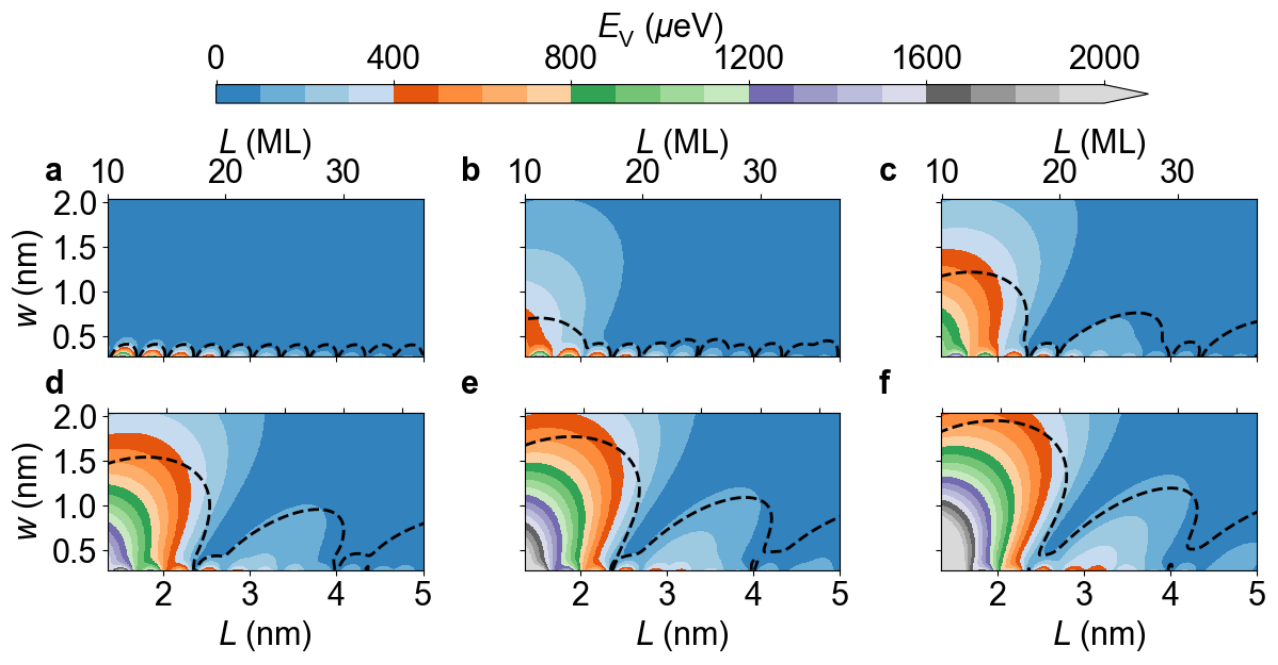}
	\centering
    \caption{
    \textbf{Evolution of the deterministic valley splitting with shear strain. }
    This figure extends the results of Figs.~3c and d in the main text to a range of shear-strain values, given by:
    \textbf{a}
    $\varepsilon_{xy}=0\%$,
    \textbf{b} $0.1\%$,
    \textbf{c} $0.2\%$,
    \textbf{d} $0.3\%$,
    \textbf{e} $0.4\%$, and 
    \textbf{f} $0.5\%$.
    Dashed lines indicate the contour $|\Delta_0|/\sigma_\Delta = \sqrt{\pi}/2$, which corresponds to the nominal onset of deterministically enhanced valley splitting.}
	\label{fig:vs_L_w}
\end{figure}

In this Supplementary Note, we study the individual contributions to the valley splitting from the conventional ($2k_0$) and the shear-strain-induced ($2k_1$) coupling mechanisms, to clarify the behavior observed in Fig.~3a of the main text. 
That figure shows the deterministic valley splitting (i.e., without disorder), as a function of interface width $w$ and shear strain $\varepsilon_{xy}$, for a quantum well of width $L=15~\mathrm{ML}=2.04$~nm. 
To distinguish the two mechanisms, we first solve for $\varphi_0(z)$ from Eq.~(4) in Methods. 
We then use this to compute the conventional $2k_0$ contribution to the valley coupling, given by
\begin{equation}
\Delta_{2k_0}=\int dz\,|\varphi_0(z)|^2V_{\mathrm{qw}}(z)e^{i2k_0z} .
\label{eq:vs_2k_0} 
\end{equation}
We also compute the shear-strain-induced $2k_1$ contribution, given by
\begin{equation}
    \Delta_{2k_1} = -i\varepsilon_{xy}\int dz\, \varphi_0^{*}(z) e^{2ik_1z} \left[C_1+C_2\left(2ik_1\partial_z+\partial_z^2\right)\right]\varphi_0(z). 
    \label{eq:vs_2k_1}
\end{equation}

We note that the quantities $\Delta_{2k_0}$ and $\Delta_{2k_1}$ are both complex, so their relative phases determine whether they interfere constructively or destructively. 
We emphasize that only the sum $\Delta_0 = \Delta_{2k_0} + \Delta_{2k_1}$ is physically accessible; however, theoretically, we may consider $\Delta_{2k_0}$ and $\Delta_{2k_1}$ separately. 
In Figs.~\ref{fig:vs_k0_k1}a and b, we plot the quantities $E_{v,2k_0}=2\left|\Delta_{2k_0}\right|$ and $E_{v,2k_1}=2\left|\Delta_{2k_1}\right|$, respectively, assuming the same simulation parameters as Fig.~3a. 
The $2k_0$ contribution (Fig.~\ref{fig:vs_k0_k1}a) is independent of the shear strain within the present model. 
In contrast, the $2k_1$ contribution (Fig.~\ref{fig:vs_k0_k1}b) is proportional to $\varepsilon_{xy}$, as anticipated from Eq.~(\ref{eq:vs_2k_1}), and vanishes when $\varepsilon_{xy} = 0$. 
The total deterministic valley splitting, shown in Fig.~3a, and reproduced in Fig.~\ref{fig:vs_k0_k1}c, is given by $E_v = 2|\Delta_0|$. 
Again, since these two contributions are complex, this splitting is not generally equal to $E_{v,2k_0}+E_{v,2k_1}$, but contains an interference term:
\begin{equation}
    E_v^2=4\big[|\Delta_{2k_0}|^2
        +|\Delta_{2k_1}|^2 +2\operatorname{Re}
        \left(\Delta_{2k_0}^*\Delta_{2k_1}\right)\big].
\label{eq:vs_interference}
\end{equation}

To investigate the onset of $2k_1$-based valley coupling, Fig.~\ref{fig:vs_L_w} shows the evolution of $E_v$ for experimentally relevant values of $\varepsilon_{xy}$, as a function of quantum-well width $L$  and interface width $w$. 
(Compare to Figs.~3c and d in the main text.)
In the absence of shear strain (Fig.~\ref{fig:vs_L_w}a), enhanced valley splittings are restricted to the ultrasharp-interface regime, defined by $w\lesssim \lambda_0 = 0.33$~nm.
Here we observe oscillations with $2k_0$ periodicity, consistent with behavior reported by Boykin \textit{et al.}~\cite{Boykin:2004p115}. 
With increasing shear strain (Figs.~\ref{fig:vs_L_w}b-f), the portions of these diagrams displaying large valley splittings grow to include a much wider range of $w$ values, and a similarly wide range of $L$ values, thereby relaxing the sharp-feature requirement associated with the conventional $2k_0$ mechanism. 
The shear-strain-induced features also show resonances with a characteristic $2k_1$ periodicity, providing an additional signature of shear-strain-induced intervalley coupling.

In each panel of Fig.~\ref{fig:vs_L_w}, the dashed black line denotes the special contour $|\Delta_0|/\sigma_\Delta = \sqrt{\pi}/2$, corresponding to the nominal threshold for deterministically enhanced behavior.
However, these contours do not precisely follow the valley splitting contours, which can be understood as follows.
Although the deterministic valley coupling $\Delta_0$ is enhanced by small $w$ or $L$ values, the standard deviation of the valley coupling $\sigma_\Delta$ reflects the electron overlap with Ge atoms, and is therefore enhanced by large $w$ or small $L$ values.
This leads to a competition between $\Delta_0$ and $\sigma_\Delta$, such that devices with large \emph{average} valley splittings can still have dangerously small valley splittings \emph{locally}.
We note that the ratio $|\Delta_0|/\sigma_\Delta$ is a much better indicator of deterministic behavior than $E_v$ \cite{Losert:2023p125405}.
Importantly, Fig.~\ref{fig:vs_L_w} shows that regions above the deterministic threshold (i.e., lying below the dashed lines) are seen to grow with increasing shear strain.

Summarizing this Note: shear strain can drive devices into the deterministically enhanced regime while also redefining what it means to be a ``narrow'' quantum well or a ``sharp'' interface.

\begin{figure}[b]
	\includegraphics[width= 18 cm]{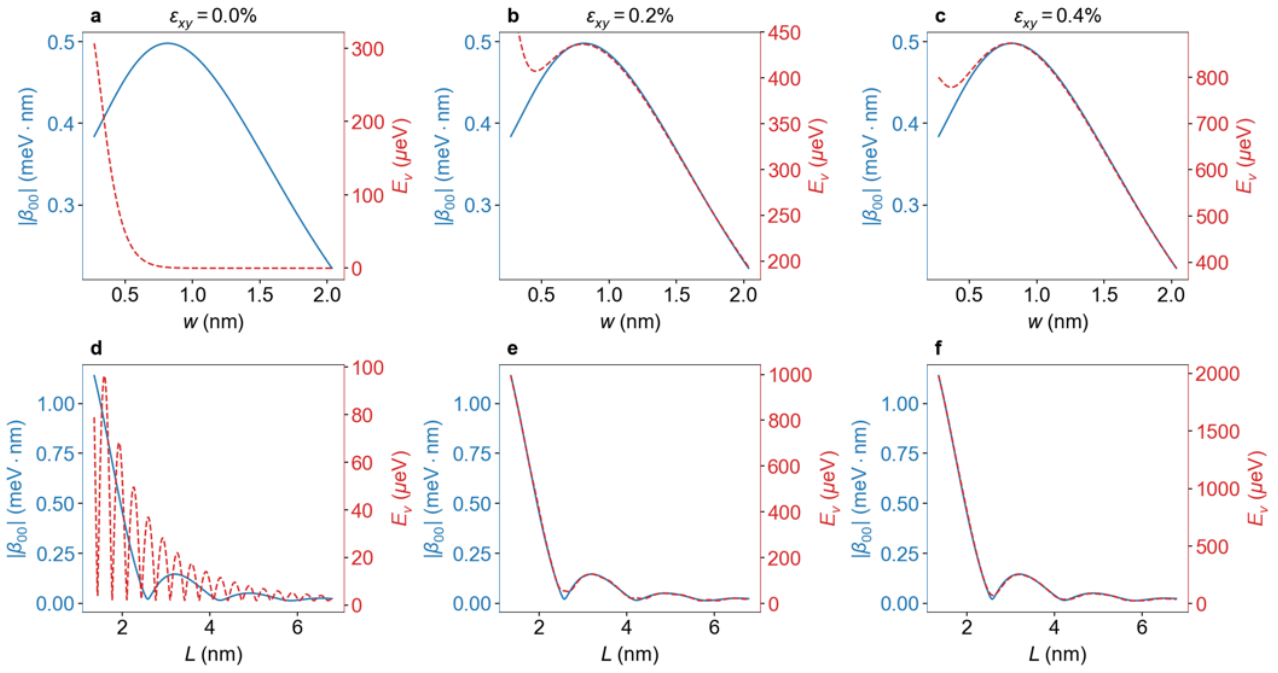}
	\centering
    \caption{\textbf{Demonstrating the approximate proportionality between the valley splitting and the spin-orbit coupling in the $2k_1$ regime.}
    The valley splitting $E_v$ (red dashed curves) and the Dresselhaus spin-orbit-coupling strength $|\beta_{00}|$ (blue solid curves) are plotted as functions of 
    \textbf{a}-\textbf{c} the interface width $w$, for a fixed quantum-well width of $L = 2$~nm, or
    \textbf{d}-\textbf{f} the quantum-well width $L$, for a fixed interface width of $w = 0.5$~nm.
    The columns correspond to shear-strain values of $\varepsilon_{xy} = 0$, 0.2, and 0.4\%, from left to right.
    We do not expect any correspondence between $E_v$ and $|\beta_{00}|$ when $\varepsilon_{xy}=0$, but we observe excellent correspondence for larger $\varepsilon_{xy}$ values, as consistent with Eq.~(\ref{eq:Evproptobeta00}).}
	\label{fig:vs_beta_proportionality}
\end{figure}

\section{Approximate proportionality of the valley splitting and spin-orbit coupling in the \texorpdfstring{$2k_1$}{2k1} regime}
Here, we examine the approximate proportionality between the valley splitting $E_v$ and the Dresselhaus spin-orbit coupling strength $|\beta_{00}|$ in the $2k_1$ regime. 
When the $\Delta_{2k_0}$ contribution [Eq.~(\ref{eq:vs_2k_0})] is negligible, the valley splitting is given by $E_v \approx 2|\Delta_{2k_1}|$. 
Moreover, we find numerically that it is dominated by the $C_1$ term in Eq.~(\ref{eq:vs_2k_1}).
The Dresselhaus coupling $\beta_{00}$, defined in Eq.~(8) of Methods, also depends on the $2k_1$ Fourier component of the electron probability density.
Thus, in the $2k_1$ regime, which we define approximately as $w> \lambda_0=0.33$~nm, we expect to observe a direct proportionality between $E_v$ and $|\beta_{00}|$, given by
\begin{equation}
E_v\approx (2C_1|\varepsilon_{xy}|/\beta_0)|\beta_{00}| .
\label{eq:Evproptobeta00}
\end{equation}

Figure~\ref{fig:vs_beta_proportionality} illustrates this relationship for the shear-strain values $\varepsilon_{xy} = 0$, $0.2\%$, and $0.4\%$ (left to right). 
Figures~\ref{fig:vs_beta_proportionality}a-c show $E_v$ and $|\beta_{00}|$ as functions of the interface width $w$, for a fixed quantum-well width of $L = 2$~nm. 
The proportionality suggested by Eq.~(\ref{eq:Evproptobeta00}) becomes quite accurate as the relative contribution from $\Delta_{2k_1}$ increases, particularly for higher shear-strain values, and when $w >\lambda_0$. 
Figures~\ref{fig:vs_beta_proportionality}d-f show $E_v$ and $|\beta_{00}|$ as functions of $L$, for a fixed interface width of $w = 0.5$~nm, again confirming the accuracy of Eq.~(\ref{eq:Evproptobeta00}) in the $k_1$ regime.

\section{Alloy disorder and statistical distributions}
The simulations reported in this work incorporate alloy disorder, which plays a key role in determining valley-coupling fluctuations and their resulting landscapes.
The theory for these disorder models has been derived elsewhere \cite{Woods:2026p206201, Losert:2023p125405, Wuetz:2022p7730}.
For completeness, we summarize the pertinent results here.

\subsection{Alloy-disorder effects in the valley coupling}
\label{supplementary: disorder}

\begin{figure}[b]
	\includegraphics[width= 18 cm]{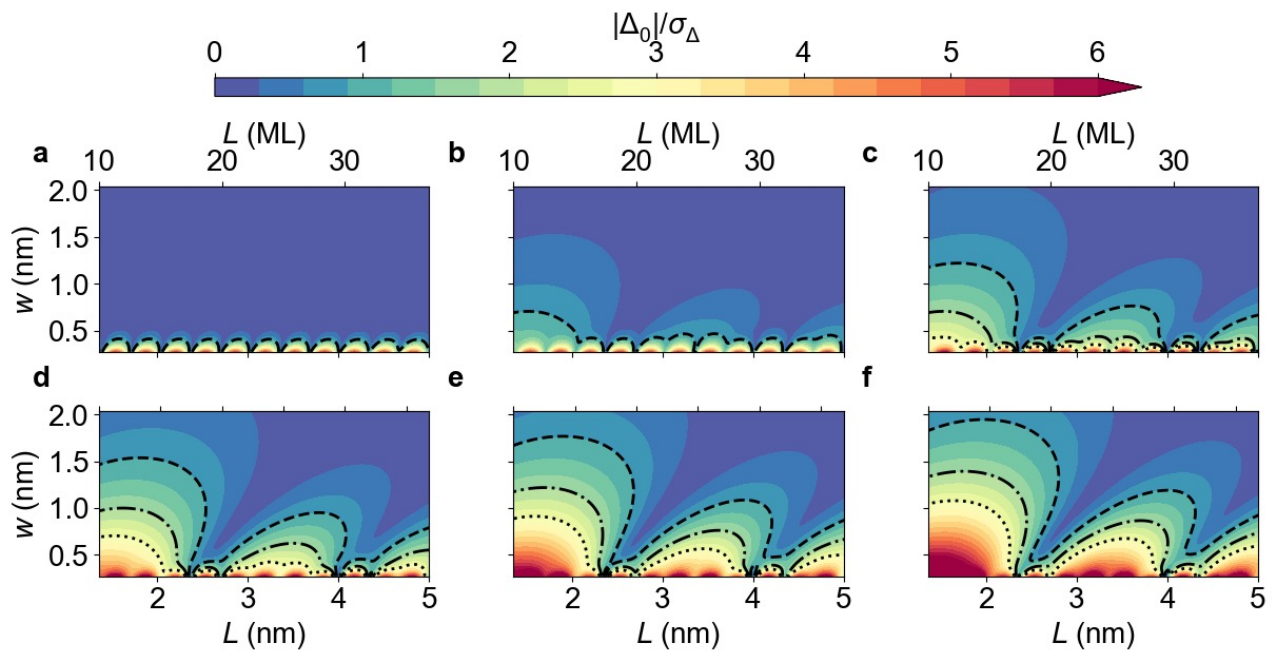}
	\centering
    \caption{
    \textbf{Evolution of the disorder parameter $|\Delta_0|/\sigma_\Delta$ with shear strain.} 
    This figure extends the results shown in Fig.~4d in the main text, for a range of shear-strain values, given by:
    \textbf{a}
    $\varepsilon_{xy}=0\%$,
    \textbf{b} $0.1\%$,
    \textbf{c} $0.2\%$,
    \textbf{d} $0.3\%$,
    \textbf{e} $0.4\%$, and 
    \textbf{f} $0.5\%$.
    For all panels: the dashed, dot-dashed, and dotted lines correspond to the contours $|\Delta_0|/\sigma_\Delta = \sqrt{\pi}/2, \sqrt{\pi},$ and $3\sqrt{\pi}/2$, respectively.
    (Only the contour $|\Delta_0|/\sigma_\Delta = \sqrt{\pi}/2$ is shown in panels \textbf{a} and \textbf{b}.)}
	\label{fig:delta_0_sigma_delta}
\end{figure}

We first characterize the variance of the valley coupling, and the crossover to deterministically enhanced behavior, using the effective-mass approach described in Refs.~\cite{Woods:2026p206201, Woods2024, Losert:2023p125405, Wuetz:2022p7730}. 
For random-alloy disorder, every site in the crystal lattice has a probability $n_{\mathrm{Ge}}$ of being occupied by a Ge atom, where $n_{\mathrm{Ge}}$ is the layer-averaged Ge concentration and $n_{\mathrm{Ge}}(z)$ describes the average quantum-well Ge profile. 
As discussed in Methods, a single Ge atom contributes an additional onsite energy of $E_{\mathrm{Ge}}=0.6~\mathrm{eV}$, compared to Si. 
It is then easy to show that the fluctuations of the potential energy at a given site have zero mean and a local variance of $E_{\mathrm{Ge}}^2 n_{\mathrm{Ge}}(z)[1-n_{\mathrm{Ge}}(z)]$.
Expressing the valley coupling as a sum of its deterministic and fluctuating components, $\Delta=\Delta_0+\Delta_\delta$, we obtain the mean value and variance of the fluctuations:
\begin{equation}
\langle\Delta_\delta\rangle=0 \quad\quad\text{and}\quad\quad
\sigma_\Delta^2 \equiv \left\langle|\Delta_\delta|^2\right\rangle
=\frac{E_{\mathrm{Ge}}^2 a_l a_t^2}{2\pi l_t^2}
\int dz\,|\varphi_0(z)|^4 n_{\mathrm{Ge}}(z)[1-n_{\mathrm{Ge}}(z)].
\label{eq:sigma_delta}
\end{equation}
Here, the ground-subband envelope function $\varphi_0(z)$ is computed using the average confinement potential, given by $E_\text{Ge}n_\text{Ge}(z)+eF_zz$ [see Eq.~(5) of Methods], which also includes an applied, vertical electric field $F_z$. 
The effective lattice constants $a_l=a/4\simeq0.136~\mathrm{nm}$ and $a_t=a/\sqrt{2}\simeq0.384~\mathrm{nm}$ reproduce the atomic density of Si, where $a$ is the cubic lattice constant. 
For the transverse confinement, we consider a simple 2D harmonic potential, with a corresponding orbital energy splitting of $\hbar\omega_0$ and an effective radius of $l_t=\sqrt{\hbar/m_t\omega_0}$, where $m_t=0.19m_e$ is the transverse effective mass.
Throughout this work, we adopt the typical value $\hbar\omega_0=1$~meV, corresponding to $l_t\simeq 20$~nm. 
We note that many aspects of the valley coupling are determined by $\sigma_\Delta$.
As seen from the previous discussion, the main experimental knobs for tuning $\sigma_\Delta$ are $n_\text{Ge}(z)$, $\hbar\omega_0$, and $F_z$.

In Figs.~4c and d of the main text, we characterize the valley-coupling variations by plotting the quantities $\sigma_\Delta$ and $|\Delta_0|/\sigma_\Delta$ as a function of the quantum-well parameters $w$ and $L$. 
In Fig.~\ref{fig:delta_0_sigma_delta}, we further explore the evolution of $|\Delta_0|/\sigma_\Delta$, for a sequence of shear-strain values.
Here, we see that the parameter range yielding deterministically enhanced behavior (lying below the dashed black lines) grows quickly with $\varepsilon_{xy}$, as expected.

\subsection{Generating spatial maps of the valley splitting}

\begin{figure}[b]
	\includegraphics[width= 17 cm]{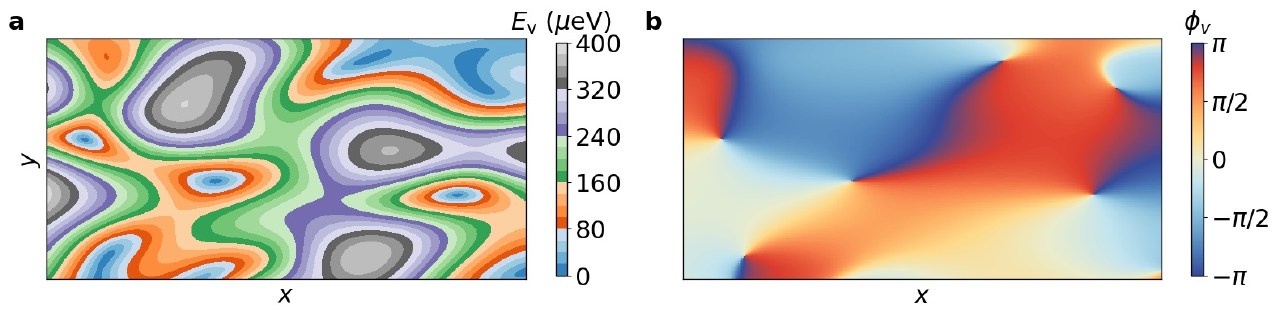}
	\centering
    \caption{
    \textbf{Spatial maps of the valley splitting and valley phase, obtained in the absence of deterministic valley coupling ($\Delta_0=0$).} 
    This figure extends the results of Figs.~5a and b in the main text, which show two different versions of a disordered valley-coupling map, for cases with nonzero deterministic valley couplings, with and without shear strain.
    \textbf{a} Here, the valley splitting originates purely from the disorder contribution: $E_v(\boldsymbol{\rho})=2|\Delta_\delta(\boldsymbol{\rho})|$. 
    \textbf{b} The corresponding disordered valley phase, $\phi_v(\boldsymbol{\rho})$.}
	\label{fig:vs_disordered}
\end{figure}

Here, we summarize the methods used to generate spatial valley-splitting maps in this work, as originally derived in Refs.~\cite{Woods:2026p206201, Losert:2023p125405, soomro2026preprint}. 
We consider isotropic, harmonically confined quantum dots of radius $l_t=\sqrt{\hbar/(m_t\omega_0)}$, as described above.
The dot centers are located at $\boldsymbol{\rho}=(x,y)$, and we consider a 2D grid of dot positions $\{\boldsymbol{\rho}_i\}$, where $i=1,\dots,N$.
Following \cite{Woods:2026p206201}, we decompose the fluctuating component of the valley coupling into its real and imaginary parts: $\Delta_\delta(\boldsymbol{\rho})=R_\delta(\boldsymbol{\rho})+iI_\delta(\boldsymbol{\rho})$.
It can be shown that the covariance matrix describing $R_\delta(\boldsymbol{\rho})$ and $I_\delta(\boldsymbol{\rho})$ is given by
\begin{equation}
\Sigma_{ij}
=\left\langle R_\delta(\boldsymbol{\rho}_i)R_\delta(\boldsymbol{\rho}_j)\right\rangle
=\left\langle I_\delta(\boldsymbol{\rho}_i)I_\delta(\boldsymbol{\rho}_j)\right\rangle
=\frac{\sigma_\Delta^2}{2}
\exp\left[-\frac{|\boldsymbol{\rho}_i-\boldsymbol{\rho}_j|^2}{2l_t^2}\right],
\label{eq:vs_r_corr}
\end{equation}
while $\langle R_\delta(\boldsymbol{\rho}_i)I_\delta(\boldsymbol{\rho}_j)\rangle=0$. 

To generate a realization of a spatial map $\Delta_\delta(\boldsymbol{\rho}_i)$, we first diagonalize the covariance tensor as
\begin{equation}
\Sigma=U\Lambda U^{\mathsf T}, \qquad
\Lambda=\operatorname{diag}(\lambda_1,\ldots,\lambda_N),
\end{equation}
where $\{\lambda_j\}$ are the eigenvalues of $\Sigma_{ij}$ and the columns of $U$ are its orthonormal eigenvectors. 
Next, we generate two vectors $\boldsymbol{\xi}_R$ and $\boldsymbol{\xi}_I$, whose entries are independent, and are drawn from a normalized Gaussian distribution.
Spatial maps satisfying the covariance relation of Eq.~(\ref{eq:vs_r_corr}) are then given by
\begin{equation}
\boldsymbol{R}_\delta=U\Lambda^{1/2}\boldsymbol{\xi}_R
\quad\quad\text{and}\quad\quad
\boldsymbol{I}_\delta=U\Lambda^{1/2}\boldsymbol{\xi}_I .
\end{equation}

In Figs.~5a and b of the main text, we plot two different versions of a disordered valley-splitting map, corresponding to cases with and without shear strain, respectively.
These two plots differ in several ways, including the fact that they have very different deterministic valley-coupling components. 
Due to the absence of shear strain in Fig.~5a, the deterministic component is rather small: $|\Delta_0|= 11.7$~$\mu$eV, arising from the confinement potential.
In Fig.~5b, the shear strain is large and spatially varying, yielding a spatially varying $\Delta_0(x,y)$ map with a larger average value.
On the other hand, the valley-coupling fluctuations, $\Delta_\delta(x,y)$, are determined by the alloy disorder, and are therefore the same for both cases.
In Fig.~\ref{fig:vs_disordered}, we plot maps obtained for just the $\Delta_\delta(x,y)$ component of the valley coupling, for the same disorder realization used in Fig.~5, with the deterministic component removed (i.e., $\Delta_0=0$).
Here, we use the values $l_t\simeq20~\mathrm{nm}$ and $\sigma_\Delta=118~\mu\mathrm{eV}$.
Figure~\ref{fig:vs_disordered}a shows the resulting valley-splitting map, $E_v(\boldsymbol{\rho})=2|\Delta_\delta(\boldsymbol{\rho})|$, while Fig.~\ref{fig:vs_disordered}b describes the corresponding valley phase, $\phi_v(\boldsymbol{\rho})$.

\section{Statistics of the valley splitting, valley phase, and \texorpdfstring{$g$}{g}-factor}
\label{sec:distribution_functions}
In this Note, we summarize or derive the probability distribution functions for three different quantities of interest.

As discussed above, the real and imaginary parts of $\Delta_\delta$ are uncorrelated Gaussian random variables, with zero mean, and a variance of $\sigma_\Delta^2/2$. The conditional probability distribution for $\Delta$ is then given by
\begin{equation}
f_\Delta(\mathrm{Re}[\Delta],\mathrm{Im}[\Delta]; |\Delta_0|,\sigma_\Delta, \phi_0)
=\frac{1}{\pi\sigma_\Delta^2}
\exp\left[
-\frac{
\left(\mathrm{Re}[\Delta]-|\Delta_0|\cos\phi_0\right)^2+
\left(\mathrm{Im}[\Delta]-|\Delta_0|\sin\phi_0\right)^2
}{\sigma_\Delta^2}
\right] ,
\label{eq:bivariate_delta_dist}
\end{equation}
where the deterministic valley phase $\phi_0$ is defined by $\Delta_0=|\Delta_0|e^{i\phi_0}$.

\begin{figure}[t]
	\includegraphics[width= 17 cm]{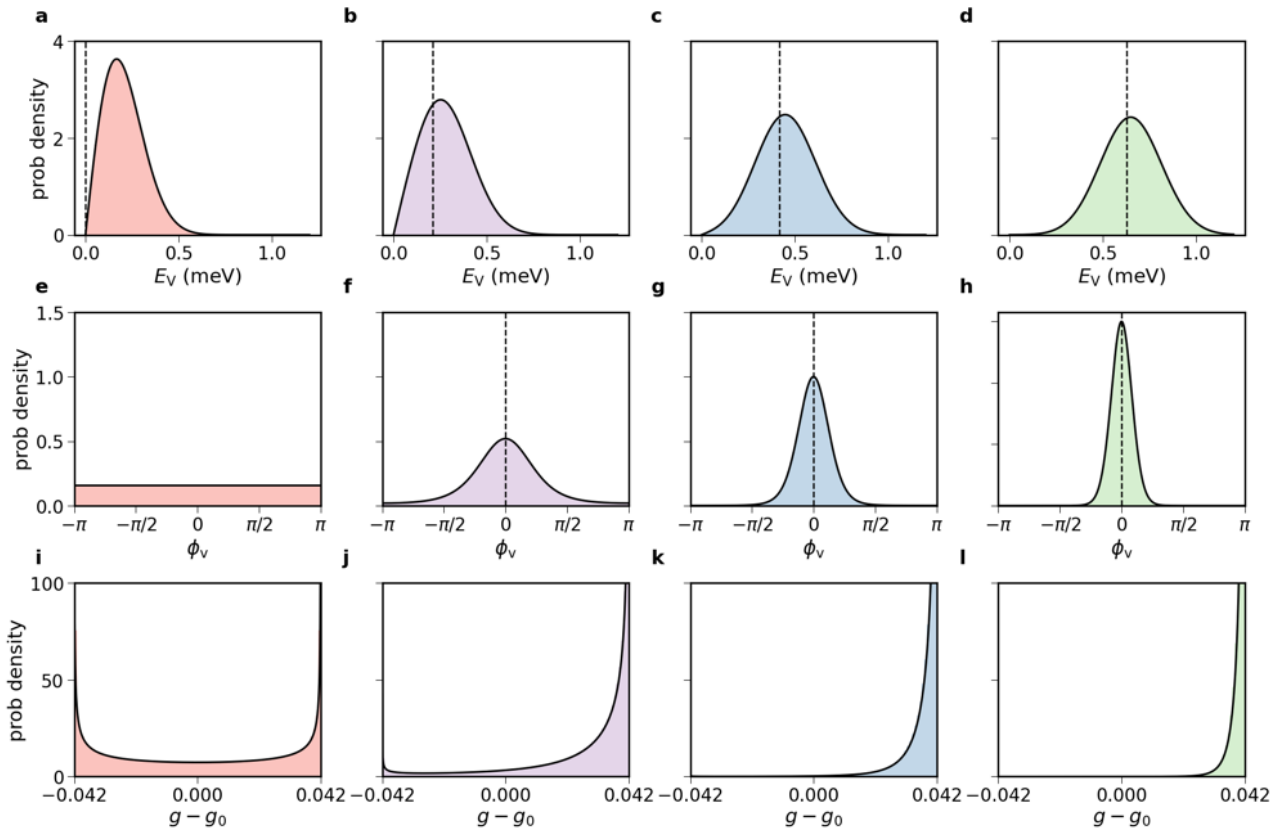}
	\centering
    \caption{
    \textbf{Evolution of three different distribution functions, for several values of the disorder parameter $|\Delta_0|/\sigma_\Delta$.}
    Since the nominal crossover to deterministic enhancement occurs when $|\Delta_0|/\sigma_\Delta > \sqrt{\pi}/2$, we consider here the cases of $|\Delta_0|/\sigma_\Delta = 0$, $\sqrt{\pi}/2$, $\sqrt{\pi}$, and $3\sqrt{\pi}/2$ (left to right), for three different distribution functions (top to bottom), where we set $\sigma_\Delta = 0.118$~meV.
    In all plots, the colored regions correspond to histograms obtained from $10^7$ independently sampled disorder realizations, assuming $\sigma_\Delta = 0.118$~meV.
    (Note that the colors used in this figure are not intended to match colors used elsewhere in this work.)
    \textbf{a}-\textbf{d} Valley-splitting distributions.
    Here, the solid-black lines correspond to the Rician distribution of Eq.~(\ref{eq:rice_dist}).
    The vertical dashed lines correspond to the deterministic valley splitting, $E_v = 2|\Delta_0|$.
    Note that, while \textbf{a} has no deterministic valley coupling ($|\Delta_0|=0$), it is still difficult to distinguish from the case of significant deterministic enhancement, corresponding to $|\Delta_0|/\sigma_\Delta=\sqrt{\pi}/2$ in \textbf{b}.
    Also note the exponential suppression of the lowest $E_v$ values in \textbf{d}.
    \textbf{e}-\textbf{h} Valley-phase distributions.
    Here, the solid-black lines correspond to Eq.~(\ref{eq:valley_phase_distribution}).
    The vertical dashed lines correspond to the deterministic valley phase, $\phi_0$.
    In this case, the deterministically enhanced distributions are much easier to verify, because they are strongly narrowed.
    \textbf{i}-\textbf{l} $g$-factor distributions.
    Here, the solid-black lines correspond to Eq.~(\ref{eq:g_factor_pdf}).}   
	\label{fig:vs_distribution}
\end{figure}

Equation (\ref{eq:bivariate_delta_dist}) describes both the amplitude and phase of the valley coupling, which we can explore separately.
For example, integrating over $\phi_0$ gives the well-known Rician distribution for the valley splitting \cite{Losert:2023p125405, woods2025preprint}, defined as $E_v=2|\Delta|$:
\begin{equation}
f_{E_v}(E_v;|\Delta_0|,\sigma_\Delta)
=\frac{E_v}{2\sigma_\Delta^2}
\exp\left[-\frac{E_v^2+4|\Delta_0|^2}{4\sigma_\Delta^2}\right]
I_0\left(\frac{E_v|\Delta_0|}{\sigma_\Delta^2}\right),
\label{eq:rice_dist}
\end{equation}
where $I_0$ is a modified Bessel function of the first kind.
We note that Eq.~(\ref{eq:rice_dist}) provides the expression used for the dashed lines in Fig.~5g of the main text.
In the disordered limit, $|\Delta_0|/\sigma_\Delta\rightarrow0$, Eq.~(\ref{eq:rice_dist}) reduces to a Rayleigh distribution.

An important criterion in many valley-splitting discussions is the crossover to deterministic behavior, which corresponds to the condition that the deterministic valley coupling exceeds the mean disordered contribution.
This criterion can be derived from Eq.~(\ref{eq:rice_dist}) by setting the deterministic component to zero ($|\Delta_0|=0$) and computing the mean value $\langle|\Delta_\delta|\rangle=\sqrt{\pi}\sigma_\Delta/2$, which leads to the desired crossover criterion: $|\Delta_0|/\sigma_\Delta>\sqrt{\pi}/2\simeq0.9$.
We note that this crossover is gradual, with deterministic behavior becoming more pronounced as the $|\Delta_0|/\sigma_\Delta$ ratio increases.
This can be seen in Figs.~\ref{fig:vs_distribution}a-d, where Rician distributions are plotted for several multiples of the crossover criterion: $|\Delta_0|/\sigma_\Delta = 0$, $\sqrt{\pi}/2$, $\sqrt{\pi}$, and $3\sqrt{\pi}/2$.
In all panels of Fig.~\ref{fig:vs_distribution}, we assume $\sigma_\Delta=0.118$~meV, which characterizes the disorder level for a quantum well with $L = 15$~ML and $w= 4$~ML, corresponding to the starred parameters in Fig.~4c of the main text.
In Figs.~\ref{fig:vs_distribution}c and d, we observe the emergence of strong deterministic enhancement, for which the distribution function is suppressed for the lowest $E_v$ values.
On the other hand, we note the similarity of the $E_v$ distributions in Figs.~\ref{fig:vs_distribution}a and b, which makes it difficult to distinguish deterministic from random behavior, for low $|\Delta_0|/\sigma_\Delta$ values.

Equation (\ref{eq:rice_dist}) also determines the distribution of the valley phase, $\phi_v$ (defined in $\Delta=|\Delta|e^{i\phi_v}$), by integrating over $|\Delta|$, giving
\begin{equation}
f_{\phi_v}(\phi_v;|\Delta_0|,\sigma_\Delta,\phi_0)=\frac{1}{2\pi}e^{-|\Delta_0|^2/\sigma_\Delta^2}+\frac{|\Delta_0|\cos(\phi_v-\phi_0)}{2\sqrt{\pi}\sigma_\Delta}e^{-|\Delta_0|^2\sin^2(\phi_v-\phi_0)/\sigma_\Delta^2}\left[1+\operatorname{erf}\left(\frac{|\Delta_0|\cos(\phi_v-\phi_0)}{\sigma_\Delta}\right)\right].
\label{eq:valley_phase_distribution}
\end{equation}
This provides the expression used for the dashed lines in Fig.~5h of the main text.
In the disordered limit, where $|\Delta_0|/\sigma_\Delta\rightarrow0$, we see that the valley phase becomes uniformly distributed across $[-\pi,\pi)$. 
However, as the deterministic contribution grows, the distribution becomes progressively more localized about $\phi_0$. 
Thus, increasing $|\Delta_0|/\sigma_\Delta$ suppresses not only fluctuations of the valley-splitting magnitude but also the valley phase.
This can be seen in Figs.~\ref{fig:vs_distribution}e-h, where we plot the valley-phase distributions corresponding to Figs.~\ref{fig:vs_distribution}a-d.
In \cite{woods2025preprint}, it was argued that, although deterministic valley coupling can be difficult to verify  in $E_v$ distributions (e.g., in Figs.~\ref{fig:vs_distribution}a and b), the situation is much improved when valley phase distributions are considered (e.g., Figs.~\ref{fig:vs_distribution}e and f).

As noted in the main text, fluctuations of the valley phase are also reflected as fluctuations in the $g$-factor. 
It is therefore important to derive the $g$-factor distribution, since its statistics are relatively easy to gather in experiments.
Setting $\theta_B=\pi/2$ in Eq.~(9), we write $g-g_0\approx |g_\tau|\cos(\phi_g-\phi_v)$ for the valley ground state.
It is convenient to redefine the phase variables as $\phi_v' = \phi_g-\phi_v$ and $\phi_0'=\phi_g-\phi_0$, such that $g-g_0=|g_\tau|\cos\phi_v'$. 
The distribution of $\phi_v'$ has the same form as Eq.~\eqref{eq:valley_phase_distribution}, with $\phi_v-\phi_0$ replaced by $\phi_v'-\phi_0'$. 
We note that all valley phase quantities in this work, including spatial maps and statistical distributions, have been plotted using these primed variables, defined relative to $\phi_g$.
The probability distribution for the variable $x\equiv g-g_0=|g_\tau|\cos\phi_v'$ can then be solved using standard formulas for transformations of random variables \cite{Casella2002}. 
Here we note that, since the cosine function is two-to-one over the interval $[-\pi,\pi)$, both phase branches must be considered. 
Defining the variable $\theta_x=\cos^{-1}\left(x/|g_\tau|\right)$, we finally obtain
\begin{equation}
f_g(x;|\Delta_0|,\sigma_\Delta,\phi_0)=\frac{f_{\phi_v'}(\theta_x)+f_{\phi_v'}(2\pi-\theta_x)}{|g_\tau|\sqrt{1-\left(x/|g_\tau|\right)^2}},
\qquad \text{where}\quad
-|g_\tau|<x<|g_\tau|.
\label{eq:g_factor_pdf}
\end{equation}
This provides the expression used for the dashed lines in Fig.~5i of the main text.
It is clear that the statistical properties of the $g$-factor are inherited directly from those of the valley phase. 
This can be seen in Figs.~\ref{fig:vs_distribution}i-l, where we plot the $g$-factor distributions corresponding to Figs.~\ref{fig:vs_distribution}e-h, observing similar narrowing of the distributions.

\end{document}